\documentstyle[floats,aps]{revtex}
\input amssym.def
\input amssym.tex

\input epsf
\begin{document}

\draft
\twocolumn[\hsize\textwidth\columnwidth\hsize\csname
@twocolumnfalse\endcsname

\title{Point-Contact Conductances at the Quantum Hall Transition}

\author{Martin Janssen$^1$, Marcus Metzler$^{1,2}$, and Martin R. Zirnbauer$^1$
 \\\ $^1$ Institut f\"ur Theoretische Physik, Universit\"at zu K\"oln,
  \\\ Z\"ulpicher Str. 77, 50937 K\"oln, Germany
\\\ $^2$ Department of Physics, Toho University, \\\ Miyama 2-2-1,
Funabashi, Chiba 274-8510}

 \date{February 1999 (revised version)}
\maketitle

\begin{abstract}
  On the basis of the Chalker-Coddington network model, a numerical
  and analytical study is made of the statistics of point-contact
  conductances for systems in the integer quantum Hall regime.  In the
  Hall plateau region the point-contact conductances reflect strong
  localization of the electrons, while near the plateau transition
  they exhibit strong mesoscopic fluctuations.  By mapping the network
  model on a supersymmetric vertex model with ${\rm GL}(2|2)$
  symmetry, and postulating a two-point correlator in keeping with the
  rules of conformal field theory, we derive an explicit expression
  for the distribution of conductances at criticality.  There is only
  one free parameter, the power law exponent of the typical
  conductance.  Its value is computed numerically to be $X_{\rm t} =
  0.640 \pm 0.009$.  The predicted conductance distribution agrees well
  with the numerical data.  For large distances between the two
  contacts, the distribution can be described by a multifractal
  spectrum solely determined by $X_{\rm t}$.  Our results demonstrate
  that multifractality can show up in appropriate transport
  experiments.
\end{abstract}

 \vskip 2pc]

\section{Introduction}

Models of two-dimensional (2D) noninteracting electrons subject to
disorder and a strong magnetic field, form what is called the
(integer) quantum Hall universality class.  Their most prominent
feature is the existence of a localization-delocalization (LD)
transition, which underlies the plateau-to-plateau transition of the
Hall conductance observed in the integer quantum Hall effect.  Among
the various members of the quantum Hall universality class, the
Chalker-Coddington network model \cite{Cha88} has been found
\cite{Kiv92Fish96NN96,Kle95} to be a convenient representative,
particularly for numerical purposes.  A wavefunction in this model is
a collection of complex amplitudes, one for each bond of a square
lattice.  The time evolution operator acts on the wavefunctions by
discrete steps, which are determined by unitary scattering matrices
assigned to the vertices of the lattice.  It has been established
that, with varying left-right asymmetry of the scattering probability
$p \in [0,1]$, the stationary states undergo a LD transition
\cite{Kiv92Fish96NN96} with a critical exponent $\nu \approx 2.35$ for
the localization length: $\xi \sim |p - p^{\ast}|^{-\nu}$, $p^{\ast} =
0.5$.  Moreover, it was shown that the critical states (for which the
localization length is much larger than the system size $L$) have
multifractal properties \cite{Kle95} that are universal, and are
characterized by an exponent $\alpha_0 \approx 2.28$ describing the
scaling of the typical value (i.e. the geometric mean) of the squared
amplitude: $\exp\left\langle\ln|\psi|^2\right\rangle \sim
L^{-\alpha_0}$.  The combination $\nu(\alpha_0-d)$ is the critical
exponent of the typical local density of states (LDoS), which has been
argued \cite{JanR97} to be an order parameter for the LD transition.

The critical exponent $\nu$ was extracted in a number of transport
experiments \cite{exp}.  In contrast, no such experiment has been
carried out to measure $\alpha_0$ or related multifractal exponents.
Multifractality is observable in local quantities such as local
densities, or in the wave number and frequency dependent dynamic
structure factor (cf.~\cite{ChaDan88,KleHuc97}).  These have yet to be
studied in transport measurements under mesoscopic conditions.  In
\cite{Fas92} it was pointed out that multifractality can show up in
the size dependence of the conductance distribution close to the LD
transition.  Further suggestions for an experimental determination
were made in \cite{ShiBra}, but it seems that these are still awaiting
realization.  Recently, it has been suggested that multifractality
relates to the corrections to scaling and may be observable in the
temperature dependence of the peak-hight of the conductance at the LD
transition \cite{Pol98}.  In the present paper we demonstrate that
another sensitive probe are {\it point-contact conductances}, making
multifractality directly accessible through a suitable transport
measurement. By a point-contact conductance we mean a conductance
between two small interior probes separated by a distance $r$. They
show strong mesoscopic fluctuations at the LD transition, similar to
those of the conductance (see e.g. \cite{Cob96,WanChoSou,Xio97,Gal}).
By varying the distance $r$, point-contact conductances allow to study
local details of mesoscopic fluctuations that are not captured by the
(global) conductance.

One of our motivations came from Ref.~\cite{mrz_iqhe}, which discusses
the conductance between two (or more) small interior contacts from a
field theoretic perspective.  In that work, it was pointed out that
for a large enough system the point-contact conductance depends only
on the distance $r$ between the interior contacts, whereas the typical
LDoS also involves the system size $L$.  At criticality, the
conductance in the infinite plane falls off algebraically with $r$
and, by the conformal hypothesis, this decay should be conformally
related to the decay in other geometries such as the cylinder.
Unfortunately, a comprehensive analytical theory of the critical point
does not yet exist, and we cannot predict the critical exponents.  As
we are going to show, however, we are able to draw some strong
conclusions just from the assumption of the existence of a conformal
field theory for the critical point.

In the present work, we use the Chalker-Coddington model to calculate
point-contact conductances as well as quasi energies and the
corresponding stationary states.  In Sec.~\ref{sec:LDoS} we consider
the dynamics of a closed network. To begin with familiar objects, we
demonstrate that, under critical conditions, the local density of
states has multifractal correlation exponents that agree with those
obtained from Hamiltonian models \cite{Pra96}. These results will
later be contrasted with the critical behavior of point-contact
conductances in Sec.~\ref{sec:g-crit}.  We then review the definition
of the point-contact conductances and compute them in the localized
regime. We find them to be well described by a log-normal distribution
determined by a single parameter, the typical localization length
$\xi_{\rm t}$ (Sec.~\ref{sec:g-loc}).  Our main theme, the
investigation of point-contact conductances at $p^\ast = 1/2$, is
taken up in Sec.~\ref{sec:g-crit}, which employs a combination of
analytical and numerical techniques, and is divided into five
subsections.  The $q^{\rm th}$ moment of the conductance, $\langle
T^q\rangle$, is expressed as a two-point correlator of a
supersymmetric vertex model equivalent to the network model
\cite{mrz_iqhe,mrz_network,Gru97}.  We exploit the global ${\rm
  GL}(2|2)$ symmetry of that model and, based on the general
principles of conformal field theory, propose a closed analytical
expression for $\langle T^q\rangle$.  From that we extract predictions
for the typical conductance, the log-variance and the entire
distribution function.  These predictions leave one parameter
undetermined, which can be identified with the power law exponent
$X_{\rm t}$ of the typical conductance.  We calculate $X_{\rm t}$
numerically and find $X_{\rm t} = 0.640 \pm 0.009$.  The distribution
function with this value is shown to agree with the numerical data.
In the limit of large separation between the contacts, which is
difficult to reach numerically, the conductance distribution can be
reduced to a spectrum of multifractal exponents.  We speculate on a
possible connection between the multifractal spectra of the
point-contact conductances and of the local density of states.

Finally, by matching to results that are available for the quasi-1D
limit, we argue that, if the network model (or a suitable continuum
limit thereof) were a fixed point of the renormalization group, then
the power law exponent of the typical conductance would have to be
$X_{\rm t} = 2/\pi \approx 0.637$, which is in remarkably close
agreement with our result from numerics.  Because standard conformal
field theories of the Wess-Zumino-Witten or coset type predict
critical exponents to be rational numbers (if the fixed point is
isolated) such a value, if correct, would imply an unconventional
fixed point theory.

\section{Multifractality of the local density of states at criticality}
\label{sec:LDoS}

To describe the critical behavior of 2D electrons in a strong magnetic
field and a smooth random potential, Chalker and Coddington
\cite{Cha88} formulated a network model which is very simple and yet
captures the essential features.  The model is composed of a set of
elementary ``scatterers'' placed on the vertices of a square lattice.
(In a microscopic picture, these correspond to the saddle points of
the random potential \cite{Fer88}.)

\begin{figure} [ht]
\leavevmode
\epsfxsize=6cm
\epsfysize=6cm
\centerline{\epsfbox{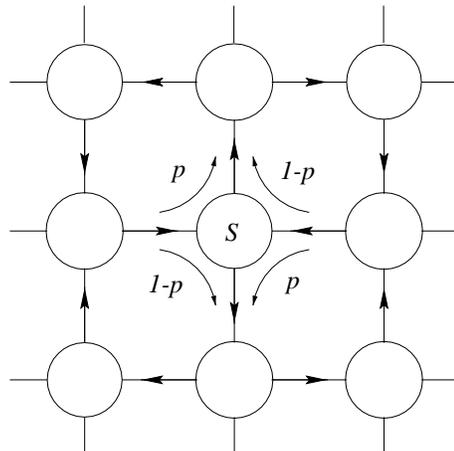}}
\caption{\label{FIG1}  Graphical representation of the 
  Chalker-Codding\-ton network model. Wave amplitudes propagating on the
 links are scattered to the left (right) with probability $p$ (resp.
  $1-p$).  The corresponding unitary scattering matrices $S$ are
  situated at the nodes of the network.}
\end{figure}

Unidirectional channels link the scatterers to each other as shown in
Fig.~1.  Each elementary scatterer is represented by a unitary
$2\times 2$ scattering matrix $S$ transforming $2$ incoming amplitudes
into $2$ outgoing amplitudes.  To simulate the effect of a disordered
or irregular array of scatterers, a (kinetic) phase factor is attached
to every channel amplitude.  These phases are taken to be independent
random variables distributed uniformly on the interval $[0,2\pi]$.  An
incoming amplitude on a given link can only be scattered to the left
or right.  Let us denote the probability for scattering to the left by
$p$.  The scattering probability to the right is then $1-p$.  The
parameter $p$ for each scatterer can be taken to be fixed or drawn at
random from a certain distribution.  In either case, the states in a
system of finite size $L$ turn out to be delocalized
\cite{Cha88,Kiv92Fish96NN96} when the mean of $p$ lies in a small
interval around $p = 0.5$.  The width of this interval, $\Delta p$,
shrinks to zero in the thermodynamic limit, $\Delta p \sim
L^{-1/\nu}$, where $\nu \approx 2.35$ is the critical exponent of the
localization length $\xi$.
 
A network model wavefunction is a set of complex amplitudes $\left
\lbrace\psi(l) \right\rbrace$ where $l = 1, ..., N_l$ runs over the
links of the network, and the normalization is fixed by $\sum_l
|\psi(l)|^2 = 1$.  Wavefunctions are propagated forward in time by
discrete steps \cite{Edr89,Kle95},
\begin{equation}
  \psi_{t+1}(l) = \sum_{l^\prime = 1}^{N_l} 
  U(l,l^\prime) \psi_t(l^\prime) \;,
  \label{3.70}
\end{equation}
the propagator for one unit of time being a sparse unitary $N_l \times
N_l$ matrix $U$ which is uniquely determined by the network of
scattering matrices \cite{Note4}.

Stationary states of the (isolated) network are solutions of the
equation \cite{Kle95,Kle97}
  $$
  U \psi_n = {\rm e}^{i\phi_n} \psi_n \,.
  $$ 
The eigenphases $\phi_n$ will be referred to as {\it quasi energies}. 
(This terminology is motivated by the analogy with the eigenphases of
the Floquet operator of a periodically driven quantum system.)  The
stationary states of the network model are critical (which is to say
their localization lengths $\xi$ exceed the system size $L$) when the
parameter $p$ is close enough to $p^* = 0.5$.  The critical states are
multifractals \cite{Kle95}.  Criticality is visible through power law
scaling of the moments \cite{Note5}: $\left\langle |\psi(l)|^{2q}
\right\rangle_L \sim L^{-d - \tau(q)}$ where $\tau(q)$ is a nonlinear
function of $q$, and $d = 2$.  (We will continue to write $d$ instead
of $2$ to keep the dependence on the number of dimensions explicit.)
Within our numerical accuracy, the $\tau(q)$ functions for different
critical states coincide.  The corresponding distribution function is
specified by a single-humped positive function $f(\alpha)$, called the
multifractal spectrum of the wavefunction \cite{Kle95},
  $$
  {\rm prob}(P=|\psi(l)|^2;L) {\rm d} P \sim  
  L^{-d+f(\alpha)}{\rm d}\alpha \,,
  $$
where $\ln P \equiv -\alpha \ln L$, and $f(\alpha)$ is related to
$\tau(q)$ by a Legendre transformation $f(\alpha(q)) = q \alpha(q)
-\tau(q)$, $\alpha(q) = d\tau(q)/dq$.  In the vicinity of its maximum,
$f(\alpha)$ is well approximated by a parabola \cite{Note1}:
  $$
  f(\alpha) \approx
  d - {(\alpha-\alpha_0)^2 \over 4(\alpha_0-d)} \,,
  $$
which is seen to be determined by a single number $\alpha_0$.  From
\cite{Pra96,Kle95} we know this number to be $\alpha_0\approx 2.28$.

The local density of states (LDoS) is defined as $\rho(\phi,l) =
\sum_{n} \delta(\phi-\phi_n) |\psi_{n}(l)|^2 $ where $\psi_n$ denotes
a stationary state with quasi energy $\phi_n$.  To wash out the
$\delta$-peak structure of this function in a closed finite system, we
smoothen it over a scale of one mean level spacing, $\delta$.  The
LDoS then becomes
  $$
  \rho(\phi,l) = \delta^{-1} |\psi(\phi,l)|^2 \;,
  $$
where $|\psi(\phi,l)|^2$ means the square of the wavefunction
amplitude, microcanonically averaged over the quasi energy interval
$[\phi- \delta/2 , \phi+\delta/2]$.  Given $\delta^{-1} \sim L^d$
and the multifractal scaling law for the critical states, the LDoS
must scale as $\left\langle \rho(\phi,l)^q \right\rangle_L \sim
L^{- \Delta_ \rho(q)}$ with
\begin{equation}
  \Delta_\rho (q) = (1-q)d + \tau(q) \;, 
  \label{RHO}
\end{equation}
and the typical value as $\rho_{\rm t} = \exp \left\langle \ln\rho(l) 
\right\rangle_L \sim L^{d-\alpha_0}$.

The {\it average} LDoS is well known to be noncritical and
nonvanishing at the LD transition.  In contrast, the {\it typical}
LDoS does show critical behavior.  For one thing, it is zero in the
region of localized states.  For another, suppose the system under
consideration had a finite band of metallic states (which it does
not), as is the case for a mobility edge in three dimensions.  The
typical LDoS would then be nonzero in that metallic band and, for
values of $L$ much larger than the correlation length $\xi_c$, would
vanish with exponent $\beta_{\rm t}=\nu(\alpha_0-d)$ on approaching
the critical point.  Such behavior is reminiscent of an order
parameter, which distinguishes between two phases joined by a second
order phase transition.  The LDoS has in fact been proposed as an
order parameter field for the general class of LD transitions
\cite{JanR97}.  Although the 2D quantum Hall class has no extended
metallic phase, the exponent $\alpha_0-2\approx 0.28 $ still controls
the finite size scaling of the LDoS close to the critical point.

It is then natural to inquire into the nature of the {\it correlation
  functions} of the LDoS at criticality.  This was done in
\cite{JanR94,Pra96} (see also \cite{Mir97}).  The critical
correlations turned out to be of the form
\begin{eqnarray}
  \rho_2^{[q]}(\omega,r=|{\bf r}|,L) &\equiv& \left\langle
  \rho(\epsilon,l)^q \rho(\epsilon+\omega,l+{\bf r})^q \right\rangle_L
  \nonumber \\ &\sim& (r/L_{\omega})^{-z(q)} L^{-\tilde{z}(q)} \;,
  \label{8.26} 
\end{eqnarray} 
where the length $L_\omega$ is defined as the linear size of a system
with level spacing $\omega$: $L_\omega = (\omega/\delta)^{-1/d} L$
\cite{ChaDan88}.  This  length scale provides a natural cutoff for the
correlations between two critical states
\cite{Cha90,Huc94,Pra96,KleHuc97}, whence (\ref{8.26}) applies to the
regime $r \ll L_\omega < L$.  For $L_\omega > L$ the dependence on
$L_\omega$ saturates and $L_\omega$ has to be replaced by $L$ in
(\ref{8.26}).  The critical exponent $z(q)$ is given by the important
scaling relation \cite{Weg85,JanR94}
\begin{equation} 
  z(q) = d + 2\tau(q) - \tau(2q) \,.  
  \label{8.27} 
\end{equation} 
A second scaling relation,
\begin{equation} 
  \tilde{z}(q) = 2(1-q)d + 2\tau(q) = 2\Delta_\rho(q) \;, 
  \label{8.27b} 
\end{equation} 
follows from (\ref{8.26}) and (\ref{8.27}) by letting $r$ and $\omega$
go to zero and then matching with $\langle \rho^{2q} \rangle_L \sim
L^{-\Delta_\rho(2q)}$.  We emphasize that the multifractal
correlations of the LDoS have a characteristic dependence on system
size \cite{Pra96}, which is encapsulated by the family of exponents
$\tilde z(q)$.

We now report on a numerical test of the scaling relation for $z(q)$,
using the Chalker-Coddington model.  Let us start by verifying that
the critical correlations are indeed cut off at the scale $L_\omega$.

\begin{figure} [ht]
\leavevmode
\epsfxsize=7.5cm
\epsfysize=7cm
\centerline{\epsfbox{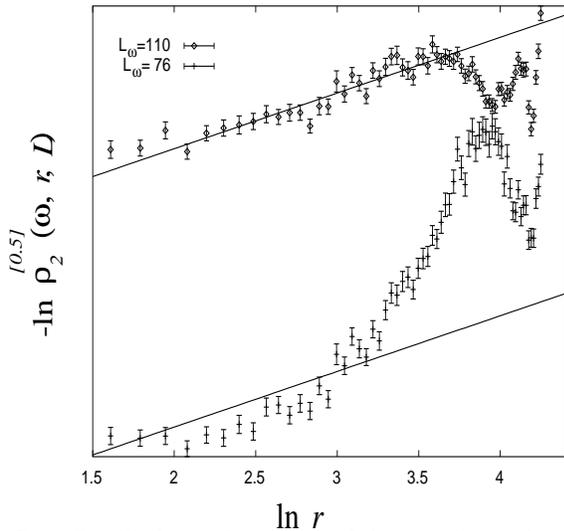}}
\caption{\label{FIG2}Double logarithmic plot of the correlation 
  function $\rho_2^{[0.5]}(\omega,r,L=150)$ versus distance $r$ for
  two different energy separations $\omega$, which correspond to
  cutoff scales $L_\omega$.}
\end{figure}

Fig.~2 plots $-\ln (\rho_2^{[q])}(\omega,r,L)$ as a function of $\ln
r$ for $q = 0.5$, $L = 150$, and two different values of $\omega$.
For $\omega = 0.000541$ (corresponding to $L_\omega \approx 76$) the
linear dependence on $\ln r$ is seen to be limited to a much smaller
region than for $\omega = 0.00026$ ($L_\omega \approx 110$).  The rest
of a our data show similar behavior, although the fluctuations due to
finiteness of the data set are quite large.  It is clear though from
these data that $L_\omega$ does set the typical scale for the cutoff
of critical correlations.

To study the scaling exponent for the $r$-dependence of the
correlator, we first took $L_\omega = \infty$, {\it i.e.} zero energy
separation, and compared the calculated exponent $z(q)$ with the value
predicted by the scaling relation (\ref{8.27}) and the $\tau(q)$
spectrum previously obtained.  As shown in the upper part of Fig.~{3},
the result for $z(q)$ agrees with the prediction within the
statistical errors.  Next, we took the quasi energies to be separated
by a finite amount corresponding to $L_\omega \approx 110$, and
calculated $z(q)$ for a few selected values of $q$.  Within the error
bars, agreement with the previous values was obtained.

\begin{figure} [ht]
\leavevmode
\epsfxsize=8cm
\epsfysize=6.5cm
\centerline{\epsfbox{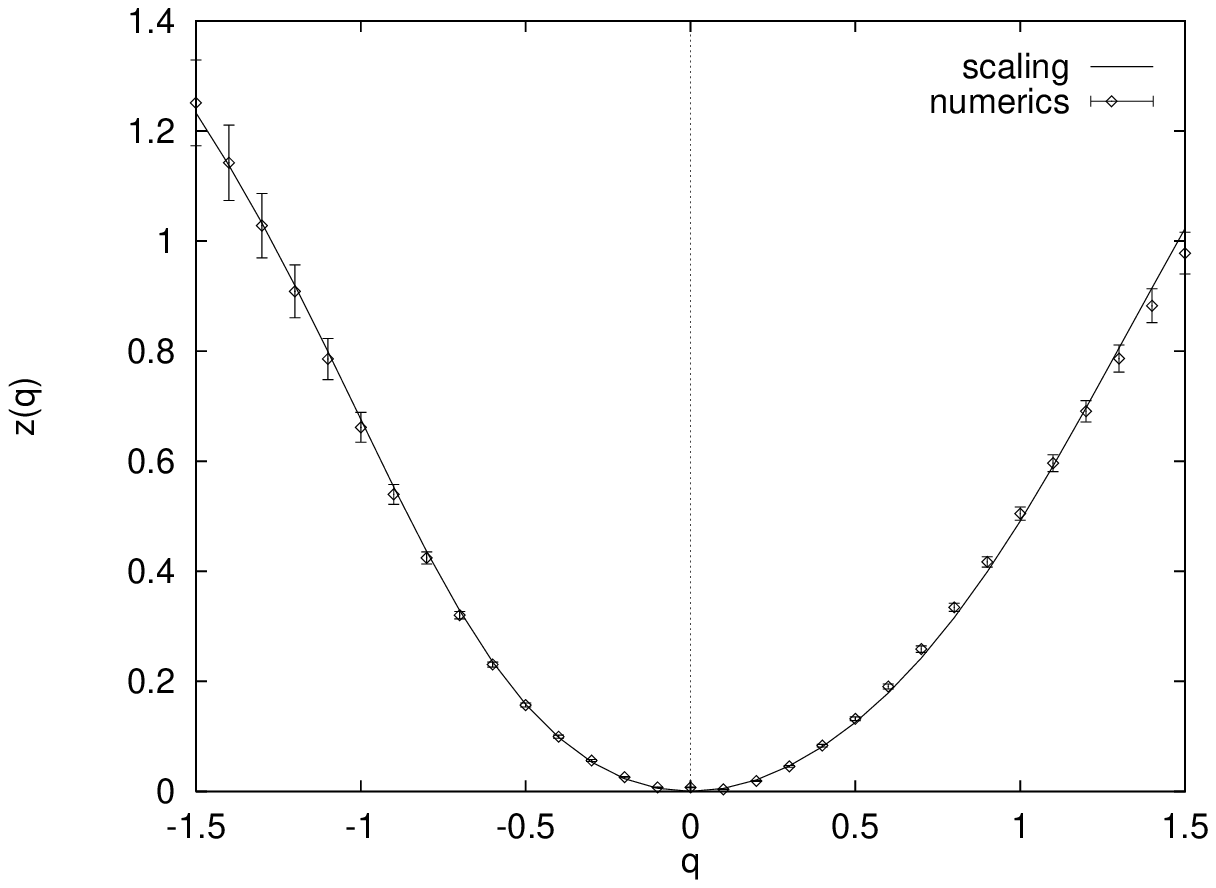}}
\epsfxsize=8cm
\epsfysize=6.5cm
\centerline{\epsfbox{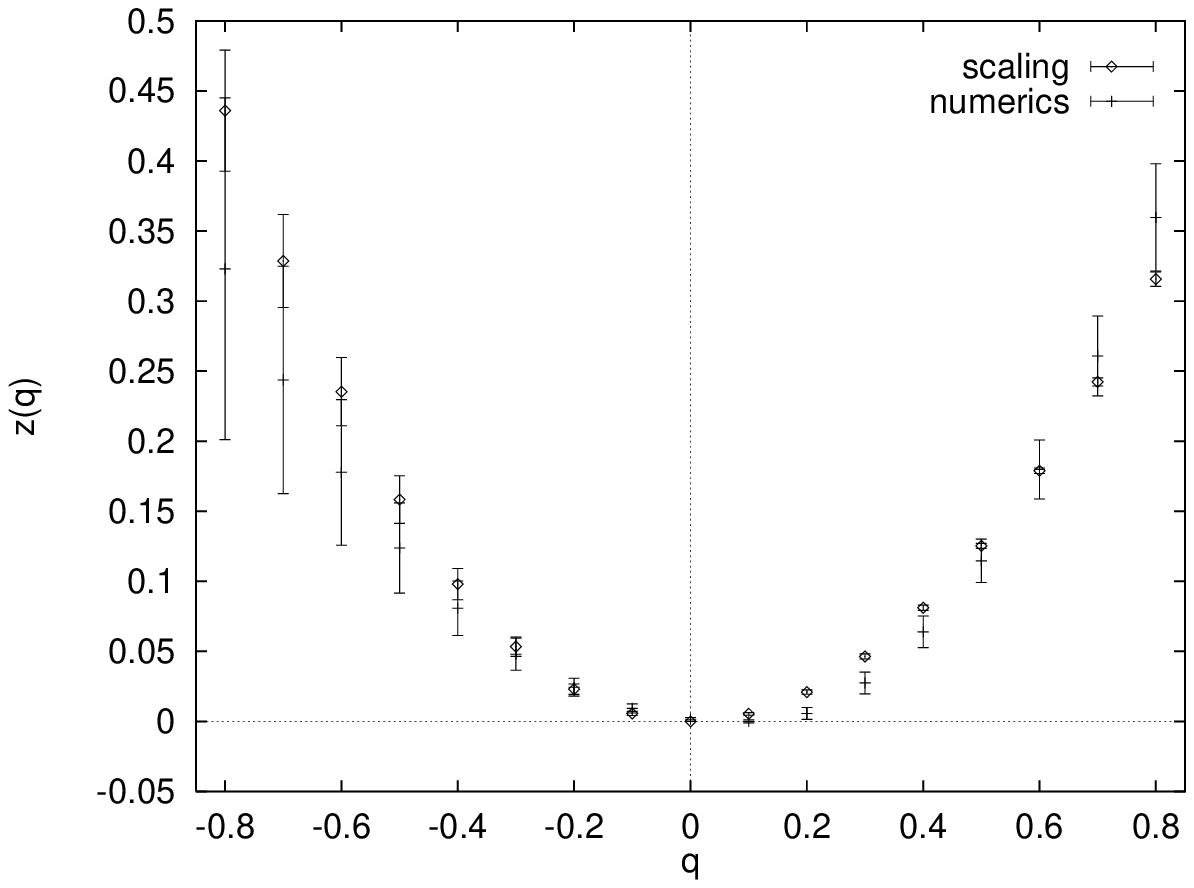}}
\caption{\label{FIG3} Scaling exponents $z(q)$ obtained from scaling
 with respect to distance $r$ (top) and energy separation $\omega$
  (bottom). The data points denoted as ``numerics'' correspond to the
 linear regression  of $\ln \rho_2^{[q]}$ versus $\ln r$
 (top) and $\ln L_\omega$ (bottom) while the data denoted as
 ``scaling'' follow from the scaling relation (\ref{8.27}) and the
  known $\tau(q)$ spectrum.}
\end{figure}

Finally, we investigated the scaling with respect to $L_\omega$ which,
according to (\ref{8.26}), should lead to the same $z(q)$.  The data
set consisted of 500 eigenfunctions for a particular disorder
realization and a range of quasi energies permitting small values of
$L_\omega$ to be reached. The system size was $L = 50$ and we
performed a spatial average.  The choice of a relatively small system
size was necessitated by the fact that the fluctuations of the
eigenfunction correlations grow in strength as $\omega$ is increased.
To reach statistical convergence, a very large number of pairings
$\rho(\phi,l)\rho(\phi+\omega,l+{\bf r})$ for fixed values of $r$ and
$\omega$ must be accumulated.  This, given present computer capacity,
is possible only for a small enough system.  We set $r = 2$ and found
behavior sufficiently linear in $\ln L_\omega$ in a regime between
$L_\omega = 3$ and $L_\omega = 12$.  The results for $z(q)$ are shown
at the bottom of Fig.~{3}.  In view of the systematic difficulties, we
conclude that our data are consistent with the scaling relation.

In summary, our numerical results for the multifractal exponents of
the local density of states clearly identify the Chalker-Coddington
model as belonging to the quantum Hall universality class.

\section{Point-contact conductances in the localized regime}
\label{sec:g-loc}

We start by reviewing what is meant by a ``point-contact conductance''
in the Chalker-Coddington model \cite{mrz_iqhe,mrz_network}.  Select
two of the links of the network for ``contacts'' and cut them in
halves (Fig.~{4}).  Then do the following.  Inject a total of one unit
of probability flux into the incoming contact links, and apply the
time evolution operator $U$ once.  Then inject another unit of flux
and apply $U$ again. Keep iterating this process, always feeding in
the same unit of probability flux so as to maintain a constant current
flow into the network.  The outgoing ends of the broken links serve as
drains, so flux will eventually start exiting through them.  After
sufficiently many iterations of the procedure, the network will have
settled down to a stationary state.  The stationary wave amplitudes at
the severed links square to transmission and reflection probabilities,
which translate into a (point-contact) conductance by the
Landauer--B\"uttiker formula.

\begin{figure} [ht]
\leavevmode
\epsfxsize=6cm
\epsfysize=5.5cm
\centerline{\epsfbox{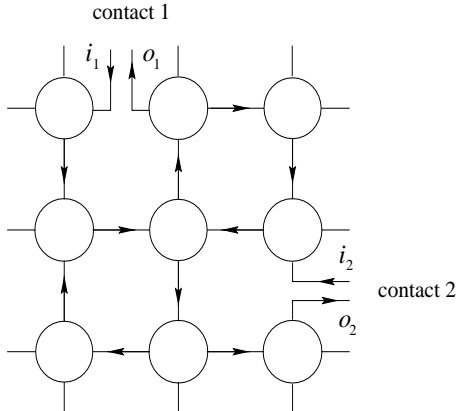}}
\caption{\label{FIG4} A section of the network model with two interior 
 point contacts.}
\end{figure}

In the present section we describe how to obtain these amplitudes by
solving a system of linear equations instead of laboring through many
iterations of the above dynamical procedure.  Afterwards, we calculate
the distribution of point-contact conductances in the localized regime
and show that it is well approximated by a one-parameter family of
functions, depending only on the value of the typical localization
length $\xi_{\rm t}$.

The injection of current into the contact links $c,c^{\prime} $ is
simulated by modifying the time step in the following way:
  $$
  |{\psi}_{t+1}\rangle = U \big( | \psi_t\rangle
   + a | c \rangle + b | {c^{\prime}} \rangle \big) \, ,
  $$
where $a$ and $b$ (subject to $|a|^2 + |b|^2 = 1$) are the amplitudes
of the current fed into the links $c$ and $c^{\prime}$, and $| c
\rangle$, $| {c^{\prime}} \rangle$ denote basis states with unit
amplitude at $c, c^{\prime} $, respectively, and zero elsewhere.  To
implement the draining action of the outgoing ends at $c$ and
$c^{\prime} $, we define projection operators $P_C$ by $P_C | \psi
\rangle = \psi(C) | C \rangle$ for $C = c, c^{\prime} $.  Using these,
we can write the complete dynamics as
\begin{eqnarray}
 | \psi_{t=0} \rangle &=& 0 \, , \nonumber \\
  | \psi_{t+1} \rangle &=& U \big( (1-P_c)
  (1-P_{c^{\prime} }) |  \psi_t \rangle
  +a | c \rangle + b | {c^{\prime}}
  \rangle \big) \, .\nonumber
\end{eqnarray}
The stationary current carrying state is formally obtained by taking
the limit $| \psi_\infty \rangle = \lim_{t\to\infty} | \psi_t
\rangle$. Alternatively, we set $U_P = U (1-P_c-P_{c^{\prime} })$ and
use stationarity to deduce for $| \psi_\infty \rangle$ the linear
equation
\begin{equation}
  \left( 1 - U_P \right) | \psi_\infty \rangle = 
  U \big( a | c \rangle + b | {c^{\prime}} \rangle  \big) \, ,
  \label{stat}
\end{equation}
which can be solved by an inversion routine.

What we want are the amplitudes of the stationary state at the links
$c$ and $c^{\prime} $, which are the components $\psi_\infty(c)$ and
$\psi_\infty(c^{\prime} )$ of the vector $| \psi_\infty \rangle$.
According to the Landauer--B\"uttiker formula, the conductance $g$
between two point contacts $c$ and $c^{\prime} $ is given by the
transmission probability $T = |t_{c^{\prime}c}|^2$ as $g = (e^2/h) T$.
The scattering problem for this is defined by feeding one unit of
current into the link $c$ and zero current into $c^{\prime}$.  The
transmission amplitude is the amplitude of $| \psi_\infty \rangle$ at
the exit $c^{\prime}$.  Hence we set $a = 1$ and $b = 0$, and compute
the conductance from
\begin{equation}
  T = | \langle c^{\prime} | \psi_\infty \rangle |^2 \,.
\end{equation}
Note that such a point-contact conductance is bounded from above by
unity: $T \le 1$.

Our numerical calculations of point-contact conductances were done for
systems of size $L = 60$, $80$, and $100$, with the distance between
contacts varying from $r = 1$ to $r = L/2$.  For every distance we
calculated between $200$ and $3000$ conductances (depending on system
size).  In the following, we present results for the distribution of
the conductance for localized states ($\xi \ll L$), corresponding to
the plateau regime of the quantum Hall effect.

\begin{figure} [ht]
\leavevmode
\epsfxsize=7cm
\epsfysize=6cm
\centerline{\epsfbox{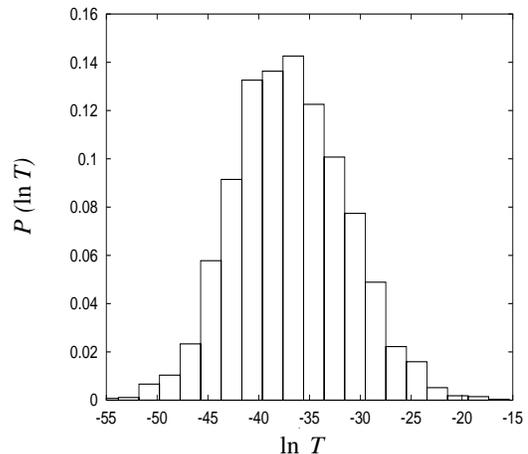}}
\caption{\label{FIG5} Normalized histogram of the logarithm of the 
  conductance $T$ in the localized regime, for a network of linear
  size $L = 60$ and a fixed distance $r = 7$ between the contacts.}
\end{figure}

Fig.~{5} shows a normalized histogram accumulated from $3000$ data
points for the logarithm of the conductance $g = T$ (in atomic units),
at fixed values of the distance $r = 7$ and the system size $L=60$.
The histogram clearly demonstrates the approximate log-normal
character of the distribution.  Such behavior is expected from the
standard picture of localization.  More precisely, the picture says
that the inverse of localization length $\xi$ has a normal
distribution (see \cite{JanR97} for a review), and the conductance is
given by $g = g_0\exp(-2 r/\xi)$ where the factor of $2$ is due to the
convention of associating the localization length with the modulus of
the wavefunction.  The inverse of the average of $1/\xi$ is called the
typical value $\xi_{\rm t}$ of the localization length: $\xi_{\rm
  t}^{-1} = \left\langle \xi^{-1} \right\rangle$.  As a further check
on this picture, we calculated $\left\langle \ln T \right \rangle$ as
a function of $r$ and found linear behavior, as shown in Fig.~{6}. The
average was performed as a combination of spatial average and disorder
average accumulating 3000 data.

\begin{figure} [ht]
\leavevmode 
\epsfxsize=7cm
\epsfysize=6cm
\centerline{\epsfbox{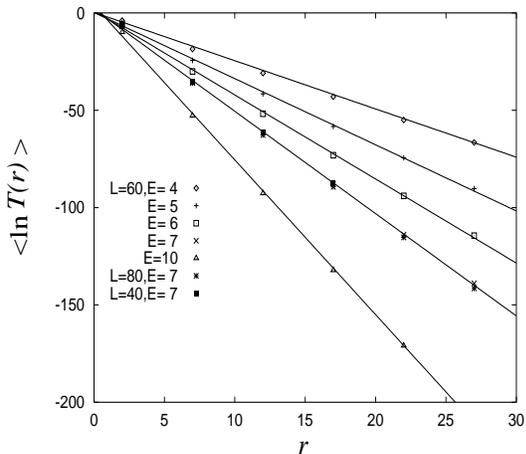}}
\caption{\label{FIG6} Plot of the average of $\ln T$ versus the
 distance $r$ between the contacts, for different system sizes $L$
 and probabilities $p = (1+{\rm e}^{-E})^{-1}$.}
\end{figure}

A more interesting question concerns the relation between the two
parameters (log-mean $\left\langle \ln T \right\rangle$ and
log-variance $\left\langle (\delta \ln T)^2 \right\rangle$) of the
log-normal distribution.  According to the one-parameter scaling
hypothesis first formulated in \cite{Abr79}, one expects the two
parameters to be dependent on each other.  Such a dependence was
indeed observed in other LD transitions (see e.g.  \cite{KraMar}), and
it takes the form of a linear relation
\begin{equation}
  \left\langle (\delta \ln T)^2 \right\rangle = 
  - A \left\langle \ln T \right\rangle + B \, ,
\end{equation}
where $A$ is a number of order unity (it equals $2$ in quasi-1D
systems \cite{ChaMac}), and $B$ is some offset due to the presence of
the factor $g_0$. Our data confirm this picture.  The value for $A$ we
find is $A = 1.00 \pm 0.05$ (Fig.~{7}).  Thus, the typical
localization length is the only relevant parameter for the conductance
distribution in the localized regime.  This distribution is very broad
and is well described by a log-normal form.

\begin{figure} [ht]
\leavevmode 
\epsfxsize=7cm
\epsfysize=5.5cm
\centerline{\epsfbox{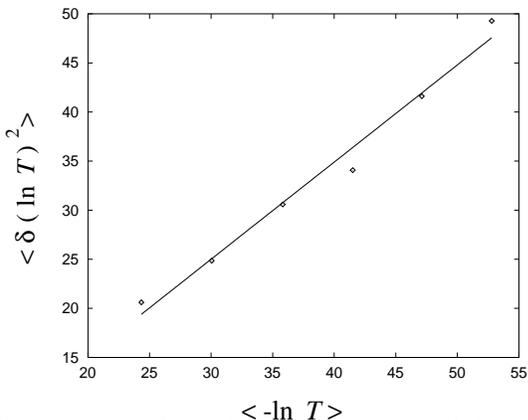}}
\caption{\label{FIG7} Linear relation between the mean and the 
 variance of the conductance distribution in the localized regime.}
\end{figure}

\section{Point-contact conductances at criticality}
\label{sec:g-crit}

We now embark on an investigation of point-contact conductances of the
critical network model at $p^\ast = 1/2$, by employing a combination
of analytical and numerical techniques.  In subsection
\ref{sec:mapping}, the network model is mapped on a supersymmetric
vertex model (for technically related work see also \cite{Gru97}) and
the $q^{\rm th}$ moment of the transmission coefficient $T$ is
expressed as a two-point correlator (Eqs.~(\ref{Tq-repr},
\ref{latticeFT})).  For this we follow the method of
\cite{mrz_network}, which is based on the so-called color-flavor
transformation and is included here for completeness.  In subsection
\ref{sec:symmetries} we take advantage of the global symmetries of the
vertex model and, based on the general principles of conformal field
theory, propose a closed analytical expression for the $q^{\rm th}$
moment of the point-contact conductance at criticality
(Eq.~(\ref{p-intq})).  This expression is analytically continued to $q
= 0$, to extract predictions for the typical conductance and the
log-variance, which are compared to numerical data in subsection
\ref{sec:typical}.  In the limit of large separation between the
contacts, which is difficult to reach numerically, a multifractal
description of conductances becomes appropriate, as is discussed in
subsection \ref{sec:multifractal}.  In subsection \ref{sec:quasi1D} we
use conformal invariance to predict an exact expression for
point-contact conductances in the quasi one-dimensional limit of the
network model.  Finally, in subsection \ref{sec:distribution} we
reconstruct from the moments the entire distribution function of
point-contact conductances.

\subsection{Mapping on a supersymmetric vertex model}
\label{sec:mapping}

We start by recalling the Landauer-B\"uttiker formula for the 
dimensionless conductance: $g = T \equiv |t_{c^\prime c}|^2$, where
  $$
  t_{c^\prime c} = \langle c^\prime | (1 - U_P)^{-1} U | c \rangle \;.
  $$
For the following it is convenient to change our conventions slightly
and interpret each of the severed contact links $c$ and $c^\prime$ as
a {\it pair} of disjoint links (Fig.~{4}), defining basis
states $| {i_1} \rangle$, $| {o_1} \rangle$ and $| {i_2} \rangle$, $|
{o_2} \rangle$ respectively.  We then impose the boundary conditions
\begin{eqnarray} 
  U | o_1 \rangle &=& 0 = U | o_2 \rangle \;,
  \label{bound_out} \\ 
  U^\dagger | i_1 \rangle &=& 0 = U^\dagger | i_2 \rangle \;.
  \label{bound_in} 
\end{eqnarray} 
The boundary conditions on the outgoing links have the same effect as
the projectors $P_c$ and $P_{c^\prime}$, which enter in the definition
$U_P = U(1 - P_c - P_{c^\prime})$.  With these boundary conditions in
force, we can write the expression for the transmission amplitude as
$t_{c^{\prime}c} = \langle o_2 | (1 - U)^{-1} U | i_1 \rangle$.
Assuming the two point contacts to be separated by more than one
lattice unit, so that $\langle o_2 | U | i_1 \rangle = 0$, we can
write it in the even simpler form
  $$
  t_{c^{\prime}c} = \langle o_2 | (1 - U)^{-1} | i_1 \rangle \;.
  $$
Note that the escape of flux, here modeled by the boundary conditions 
on the outgoing links, causes a unitarity deficit and thus ensures
positivity of the operator $1 - U^\dagger U$.

The next step is to express $t_{c^{\prime}c}$ as a Gaussian
superintegral.  For this purpose we introduce a doublet $\Psi =
(\Psi_{\rm B} , \Psi_{\rm F})$ for every link. (The incoming and
outgoing ones are included.)  The quantities $\Psi_{\rm B}$ and
$\Psi_{\rm F}$ are bosonic and fermionic ({\it i.e.} commuting and
anticommuting) complex integration variables.  Defining a Gaussian
``statistical'' average by
  $$
  \left\langle \bullet \right\rangle = \int \bullet \ \exp - S_+
  $$
with
  $$
  S_+ = \sum_{\sigma = {\rm B,F}} \sum_{l,l^\prime}
  \bar\Psi_\sigma (l^\prime) (\delta_{l^\prime l} - U_{l^\prime l}) 
  \Psi_\sigma (l) \;,
  $$
we have 
  $$
  t_{c^{\prime}c} = \left\langle \Psi_{\rm B} (o_2)
    \bar\Psi_{\rm B}(i_1) \right\rangle \;.
  $$
As usual the integration measure for the field $\Psi$ is the flat
one, normalized so that $\left\langle 1 \right\rangle = 1$.  The
Gaussian integral over $\Psi$ converges because the modulus of the
operator $U$ with boundary condition (\ref{bound_out}) is less than
unity.

Our interest here is not only in the average conductance but in {\it
  all moments} of the conductance. In fact, we ultimately will
reconstruct the entire distribution function.  To calculate the
$q^{\rm th}$ moment we need an expression for the $q^{\rm th}$ power
of $t_{c^\prime c}$. As is easily verified from Wick's theorem, this
is given by
  $$
  t_{c^{\prime}c}^q = {1 \over q!}
  \left\langle \Psi_{\rm B} (o_2)^q
    \bar\Psi_{\rm B}(i_1)^q \right\rangle \;.
  $$
Note that in a many-channel situation (such as a network model with
more than one channel per link or a conductance which is not point
contact) where $t_{c^{\prime}c}$ consists of more than one amplitude,
going from $q = 1$ to arbitrary $q$ requires enlarging $\Psi$ to a
superfield with more than one component.  Here we are in the fortunate
situation that a single component suffices to generate all the
moments \cite{Note6}.

We call $\Psi$ a ``retarded'' field and denote it from now on by
$\Psi_+$.  The complex conjugate $\overline{t_{c^{\prime}c}}$ is
expressed by a similar construction using an ``advanced'' field,
$\Psi_-$.  By combining the Gaussian integrals over retarded and
advanced fields we get 
  $$
  T^q = {1 \over q!^2} \left\langle
    \Psi_{+{\rm B}}(o_2)^q \bar\Psi_{+{\rm B}}(i_1)^q
    \Psi_{-{\rm B}}(i_1)^q \bar\Psi_{-{\rm B}}(o_2)^q
  \right\rangle
  $$
for the $q^{\rm th}$ moment of the transmission probability. The 
Gaussian statistical average here is taken with respect to $\exp -S$
where $S = S_+ + S_-$ and
  $$
  S_- = \sum_{\sigma = {\rm B,F}} \sum_{l,l^\prime} \bar\Psi_{-\sigma} (l) 
  (\delta_{ll^\prime} - \bar{U}_{l^\prime l}) \Psi_{-\sigma} (l^\prime) \;.
  $$

Our next goal is to average over the disorder of the network model.
For that we put
  $$
  U_{l'l} = U_1(l',l) \ {\rm e}^{i\varphi(l)}
  $$
where $U_1$ is the deterministic part of $U$ describing the scattering
at the nodes, and $\varphi(l)$ are random phases uniformly distributed
on the interval $[0,2\pi]$.  On extracting from $\exp -S$ the
$\varphi$-dependent parts and temporarily omitting the integration
over the superfield $\Psi$, we are faced with the integral
  \begin{eqnarray*}
    {\cal F} = \int \prod_l {d\varphi(l) \over 2\pi} 
    \exp \sum_{\sigma;l',l} \Big( \bar\Psi_{+\sigma} (l') 
      U_1(l',l) {\rm e}^{i\varphi(l)} \Psi_{+\sigma}(l&&)
    \\
    + \bar\Psi_{-\sigma} (l) {\rm e}^{-i\varphi(l)} 
    U_1^\dagger(l,l') \Psi_{-\sigma}(l') &&\Big) \;.
  \end{eqnarray*}
It is seen that the random phases ${\rm e}^{\pm i\varphi}$ couple, roughly
speaking, to the bilinears $\sum_\sigma \bar\Psi_{\pm\sigma} \Psi_{\pm
  \sigma}$.  Plain integration over the variables $\varphi(l)$ now
produces a product of Bessel functions, an expression which is not a
good starting point for further analysis.  Fortunately, there exists
something better we can do.  In the supersymmetric treatment
\cite{efetov} of disordered Hamiltonian systems, one makes a
Hubbard-Stratonovitch transformation replacing the disorder average by
an integral over a supermatrix-valued field $Q$.  It turns out that a
similar replacement, called the ``color-flavor transformation''
\cite{mrz_icmp97}, can be made in the present context.  The role of
$Q$ is taken by two sets of complex fields $l \mapsto
Z_{\sigma\sigma^\prime}(l)$ and $l \mapsto \tilde Z_{\sigma^\prime
  \sigma}(l)$, which assemble into supermatrices,
  $$
  Z = \pmatrix{Z_{\rm BB} &Z_{\rm BF}\cr Z_{\rm FB} &Z_{\rm FF} \cr}
  \;, \quad
  \tilde Z = \pmatrix{\tilde Z_{\rm BB} &\tilde Z_{\rm BF}\cr 
    \tilde Z_{\rm FB} &\tilde Z_{\rm FF} \cr} \;,
  $$
and couple to $\bar\Psi_{+\sigma} \Psi_{-\sigma^\prime}$ and $\bar
\Psi_{-\sigma^\prime} \Psi_{+\sigma}$ respectively.  The final outcome
of the transformation will be another expression for ${\cal F}$, given
at the end of the next paragraph, where instead of integrating over
the random phases $\varphi(l)$ we integrate over the supermatrix fields
$Z,\tilde Z$.

For completeness, let us now briefly summarize the mathematical
structures surrounding the supermatrices $Z$ and $\tilde Z$.  The
details can be found in \cite{mrz_circular}.  Elements $g = \pmatrix{A
  &B\cr C &D\cr}$ of the Lie supergroup $G_{\Bbb C} \equiv {\rm
  GL}(2|2)$ are taken to act on $Z$, $\tilde Z$ by the transformations
\begin{eqnarray*}
  Z \mapsto g\cdot Z &=& (AZ + B)(CZ + D)^{-1} \;,
  \\
  \tilde Z \mapsto g\cdot \tilde Z &=& (C + D\tilde Z)
  (A + B\tilde Z)^{-1} \;.
\end{eqnarray*}
Because this group action is transitive and $Z = \tilde Z = 0$ is
fixed by an $H_{\Bbb C} \equiv {\rm GL}(1|1) \times {\rm GL}(1|1)$
subgroup with elements $h = \pmatrix{A &0\cr 0 &D\cr}$, the complex
supermatrices $Z$ and $\tilde Z$ parameterize the complex coset
superspace $G_{\Bbb C} / H_{\Bbb C}$.  Let $D(Z,\tilde Z)$ be a
$G_{\Bbb C}$ invariant superintegration measure on this coset space
and put $D\mu(Z,\tilde Z) = D(Z,\tilde Z) \, {\rm SDet}(1 - \tilde Z
Z)$.  We take the integration domain for the bosonic variables to be
the Riemannian submanifold $M_{\rm r}$ defined by the conditions
 $$
 \tilde Z_{\rm FF} = - \bar Z_{\rm FF} \;, \quad
 \tilde Z_{\rm BB} = + \bar Z_{\rm BB} \;, \quad
 |Z_{\rm BB}|^2 < 1 \;.
 $$
The normalization of $D(Z,\tilde Z)$ is fixed by requiring that 
$\int D\mu(Z,\tilde Z) = 1$.  Given these definitions, it was shown 
in \cite{mrz_circular} that the above expression for ${\cal F}$ is 
equivalent to
\begin{eqnarray*}
  {\cal F} = \int {\cal D}\mu(Z,\tilde Z) \ 
  &&\exp \sum_{l \sigma \sigma^\prime} \Big( 
  \bar\Psi_{-\sigma^\prime}(l) \tilde Z_{\sigma^\prime \sigma}(l)
  \Psi_{+\sigma}(l)
  \\
  &&+ (\bar\Psi_{+\sigma} U_1)(l) Z_{\sigma\sigma^\prime}(l)
  (U_1^\dagger \Psi_{-\sigma^\prime})(l) \Big) \;,
\end{eqnarray*}
where ${\cal D}\mu(Z,\tilde Z) = \prod_l D\mu(Z(l),\tilde Z(l))$. It 
is understood that the sum over $l$ excludes all links leaving the
network.  Indeed, by the boundary condition (\ref{bound_out}) the
phase $\varphi(l)$ on an outgoing link $l = o_1$ or $l = o_2$ never
appears in the formalism and there is no need to introduce any
supermatrices $Z,\tilde Z$ there.  Alternatively, we may extend the
sum to run over all links and compensate by putting
  $$
  Z(l) = \tilde Z(l) = 0 \quad \mbox{for all outgoing links}.
  $$

We mention in passing that the equivalence between the two expressions
for ${\cal F}$ extends \cite{mrz_network} to the more general case
where the random phase factors ${\rm e}^{i\varphi(l)} \in {\rm U}(1)$ on
links are replaced by random ${\rm U}(N)$ elements.  The
transformation from these ${\rm U}(N)$ ``gauge'' degrees of freedom to
composite ``gauge singlets'' $Z(l),\tilde Z(l)$ is in striking analogy
with strongly coupled ${\rm U}(N)$ lattice quantum chromodynamics,
where integration over the gauge fields, which carry ${\rm U}(N)$
color, produces an effective description in terms of meson fields
carrying flavor.  Therefore, the passage from the first form of ${\cal
  Z}$ to the second one is referred to as the ``color-flavor
transformation'' \cite{mrz_icmp97}.

An attractive feature of the color-flavor transformation is that it
preserves the Gaussian dependence of the integrand on the fields
$\Psi, \bar\Psi$.  If we switch to schematic notation and suppress
the link and super indices, the Gaussian statistical weight is the
exponential of the quadratic form
  $$
  - ( \bar\Psi_+ \ \bar\Psi_- ) \pmatrix{1 &-U_1^{\vphantom{\dagger}} 
    Z U_1^\dagger\cr -\tilde Z &1 \cr} \pmatrix{\Psi_+\cr \Psi_-\cr} \;.
  $$
In the absence of source terms or other perturbations, integration 
over $\Psi, \bar\Psi$ simply produces the inverse of a 
superdeterminant,
  $$
  {\rm SDet}^{-1} \pmatrix{1 &-U_1^{\vphantom{dagger}} Z U_1^\dagger\cr 
    -\tilde Z &1\cr} = {\rm SDet}^{-1} 
  ( 1 - \tilde Z U_1^{\vphantom{\dagger}} Z U_1^\dagger )\;.
  $$
When the source terms are taken into account, additional factors
arise.  Since we are calculating a point-contact conductance, these
are localized at two points.  By applying Wick's theorem, we get
from the combination $\Psi_{-{\rm B}}(i_1)^q \bar \Psi_{+{\rm B}}
(i_1)^q / q!$ at the first contact an extra factor
  $$
  \left( \tilde Z (1 - U_1^{\vphantom{\dagger}} Z U_1^\dagger 
    \tilde Z)^{-1} \right)_{\rm BB} (i_1,i_1)^q = 
  \tilde Z_{\rm BB}^q (i_1)\;.
  $$
The simplification to the right-hand side of this equation occurs
because $U_1^\dagger(\bullet,i_1) = 0$ according to the boundary
conditions (\ref{bound_in}).  Unfortunately, a similar simplification
does not take place at the other contact.  (There is a basic asymmetry
between ``in'' and ``out'', which stems from the fact that we must
decide on some ordering of the factors ${\rm e}^{i\varphi}$ and $U_1$ in the
evolution operator for one time step.)  However, we can mend the
situation with a little trick. We ``prolong the exit'' by inserting an
{\it additional} node on the out-links as shown in Fig.~{8}.

\begin{figure} [ht]
\leavevmode 
\epsfxsize=4cm
\epsfysize=5cm
\centerline{\epsfbox{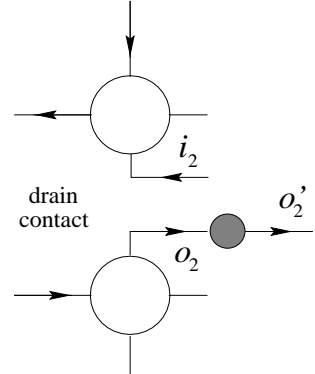}}
\caption{\label{FIG8} Prolongation of the exit by adding a dummy
  vertex at the drain contact, so as to simplify the field theoretic
  representation (see text) of the point-contact conductance.}
\end{figure}

This means that flux arriving at the link $o_2$ does not exit
immediately, but first gets transferred to $o_2^\prime$ and then exits
from there during the next time step.  With this modification, the
extra factor arising from the product $\Psi_{+{\rm B}}(o_2^\prime)^q
\bar\Psi_{- {\rm B}}(o_2^\prime)^q /q!$ simplifies to
  $$
  \left( U_1^{\vphantom{\dagger}} Z U_1^\dagger 
    (1 - \tilde Z U_1^{\vphantom{\dagger}} Z U_1^\dagger )^{-1} 
  \right)_{\rm BB} (o_2^\prime,o_2^\prime)^q = Z_{\rm BB}^q(o_2) \;,
  $$
since $U_1(o_2^\prime,o_2^{\vphantom{\prime}}) = U_1^\dagger(o_2^ 
{\vphantom{\prime}},o_2^\prime) = 1$ and $\tilde Z(o_2^\prime) = 0$. 

In summary, we have expressed the $q^{\rm th}$ moment of the 
transmission coefficient $T$ as a two-point correlator,
\begin{equation}
  \langle T^q \rangle = \left\langle Z_{\rm BB}^q (o_2) 
    \tilde Z_{\rm BB}^q (i_1) \right\rangle \;,\label{Tq-repr}
\end{equation}
of a supersymmetric lattice field theory,
\begin{equation}
  \left\langle \bullet \right\rangle = \int {\cal D}\mu(Z,\tilde Z)
  \; \bullet \; {\rm SDet}^{-1}(1 - \tilde Z U_1^{\vphantom{\dagger}} 
  Z U_1^\dagger) \;.
\label{latticeFT}
\end{equation}
The superdeterminant runs over superspace and the Hilbert space of 
the network model.  

As it stands, the formula for $\langle T^q \rangle$ does not display
clearly the internal symmetries of the theory.  To make these more
explicit, recall the definition of the (single $Z$) integration
measure,
  $$
  D\mu(Z,\tilde Z) = D(Z,\tilde Z) \ {\rm SDet}(1 - \tilde Z Z) \;.
  $$
We now factor the superdeterminant as two square roots,
  $$
  {\rm SDet}(1 - \tilde Z Z) = 
  {\rm SDet}^{1/2}(1 - \tilde Z Z) \times
  {\rm SDet}^{1/2}(1 - \tilde Z Z) \;,
  $$
and associate each of these with one of the two nodes a link begins or
ends on.  This procedure works perfectly for the internal links: every
square root factor is assigned to exactly one node.  For the external
(or contact) links the situation is different.  These are connected to
just a single node of the network, and therefore only one of the two
square roots is used up.  It is natural to assign the remaining factor
to the operator whose correlation function we are evaluating.  In this
way, the expression for $\langle T^q \rangle$ reorganizes to
\begin{equation}
  \langle T^q \rangle = 
  \left\langle v_q (o_2) v_0^\ast (i_2) \times
    v_0 (o_1) v_{-q}^\ast (i_1) \right\rangle
\label{qmoment}
\end{equation}
where
\begin{eqnarray*}
  v_q &=& Z_{\rm BB}^q \ {\rm SDet}^{1/2}(1-\tilde Z Z) \;, \\
  v_{-q}^\ast &=& \tilde Z_{\rm BB}^q \ {\rm SDet}^{1/2}(1-\tilde Z Z) \;,
\end{eqnarray*}
and
\begin{eqnarray*}
  \left\langle \bullet \right\rangle &=& \int {\cal D}(Z,\tilde Z)
  \ \bullet \ \exp - {\cal S} \;,
  \\
  {\cal D}(Z,\tilde Z) &=& \prod_l D(Z(l),\tilde Z(l)) \;.
\end{eqnarray*}
The statistical weight $\exp-{\cal S}$ is given by the superdeterminant
${\rm SDet}^{-1} (1 - \tilde Z U_1^{\vphantom{\dagger}} Z U_1 ^\dagger
)$ multiplied by one (two) factors of ${\rm SDet}(1 - \tilde Z(l)
Z(l))$ for every external (internal) link.  As was shown in
\cite{mrz_network}, $\exp -{\cal S}$ has the structure of a {\it
  vertex model} weight, {\it i.e.} it separates into a product of
factors, one for each node.  Hence the lattice field theory with
statistical weight $\exp -{\cal S}$ is called a vertex model.  The
factor for a given node, the so-called ${\cal R}$ matrix, is invariant
under the global action of ${\rm GL}(2|2)$ or, more precisely
speaking, can be regarded \cite{mrz_network} as the matrix element of
a ${\rm GL}(2|2)$ invariant operator between coherent states
parametrized by the field variables $Z,\tilde Z$ on the links which
emanate from that node.

\subsection{Symmetries and conformal hypothesis}
\label{sec:symmetries}

The special significance of the functions $v_q^{\vphantom{\ast}}$ and 
$v_{-q}^\ast$ is that they lie \cite{mrz_iqhe} in irreducible
representation spaces, denoted by $V$ and $V^\ast$, of the symmetry
group ${\rm GL}(2|2)$.  The action of ${\rm GL}(2|2)$ on these spaces
can be shown to be {\it unitary} for a ${\rm SU}(1,1) \times {\rm SU}
(2)$ subgroup, and $V$ and $V^\ast$ belong to the {\it discrete}
series of ${\rm SU}(1,1)$.  The representation space $V$ ($V^\ast$) is 
of lowest weight (resp. highest weight) type.

To make further progress, it is imperative that we exploit the global
${\rm GL}(2|2)$ symmetry of $\exp - {\cal S}$.  The links $i_1$ and
$o_1$ lie close in space, and so do $i_2$ and $o_2$, whereas these two
sets are in general far apart from each other.  Therefore, our next
step is to fuse $v_q (o_2) \in V$ with $v_0^\ast (i_2) \in V^\ast$ and
to make a decomposition into irreducible representation spaces of
${\rm GL}(2|2)$.  The operators at the other contact, $v_0 (o_1)$ and
$v_{-q}^\ast (i_1)$, are processed in the same way.  Thus we need to
know how to reduce the tensor product $V \otimes V^\ast$.  According
to Sec.~5.2 of \cite{mrz_iqhe}, this reduction involves a single
continuous series of ${\rm GL}(2|2)$, which is closely related to the
principal continuous series of unitary representations of ${\rm SU}
(1,1)$.  This series is labeled by a real parameter $\lambda \in [0,
\infty)$.  Denoting the basis state with weight $m$ by
$\varphi_{\lambda m}$, we have
\begin{eqnarray}
  v_q^{\vphantom{\ast}} v_0^\ast &=& 
  \int \left\langle V q , V^\ast 0 | \lambda q 
    \right\rangle \varphi_{\lambda q} \ \mu(\lambda) d\lambda \;,
\label{reduce1} \\
  v_0^{\vphantom{\ast}} v_{-q}^\ast &=& 
  \int \left\langle V 0 , V^\ast {\rm -}q | 
    \lambda {\rm -}q \right\rangle \varphi_{\lambda {\rm -}q} 
  \ \mu(\lambda) d\lambda \;,
\label{reduce2}
\end{eqnarray}
where $\left\langle V m , V^\ast m^\prime | \lambda m + m^\prime
\right\rangle$ is a Clebsch-Gordan coefficient, and $\mu(\lambda)
d\lambda$ is the Plancherel measure for the continuous series labeled
by $\lambda$.  Explicit expressions for these will be given below.  We
adopt the convention $\overline{\varphi_{\lambda q}} = \varphi_{
\lambda {\rm -}q}$.  Then an immediate statement is that the 
Clebsch-Gordan coefficient behaves under complex conjugation as
  $$
  \overline{ \left\langle V q , V^\ast 0 | \lambda q \right\rangle}
  = \left\langle V 0 , V^\ast {\rm -}q | \lambda {\rm -}q \right\rangle
  \;,
  $$
as follows from $\overline{v_q^{\vphantom{\ast}} v_0^\ast} = 
v_0^{\vphantom{\ast}} v_{-q}^\ast$.

All steps so far have been exact and rigorously justified.  Now we
have to make an assumption which is crucial, namely that the vertex
model at $p^\ast = 1/2$ flows under renormalization to a {\it
  conformal fixed point theory}.  Sadly, although a substantial effort
has been expended on identifying that fixed point, we still do not
understand its precise nature.  Nevertheless, as we shall see, we can
draw a number of strong conclusions just from the assumption of its
existence.

By the principles of conformal field theory, the two-point function of
$\varphi_\lambda$ decays algebraically,
\begin{equation}
  \left\langle \varphi_{\lambda q} ({\bf r}) 
    \varphi_{\lambda^\prime {\rm -}q}({\bf r}^\prime) \right\rangle = 
    {\delta(\lambda-\lambda^\prime) \over \mu(\lambda)}
    | {\bf r} - {\bf r}^\prime |^{-2 \Delta_\lambda} + \ldots
\label{twopoint}
\end{equation}
where $\Delta_\lambda$ is the scaling dimension of the most relevant
conformal field contained in the expansion of $\varphi_{\lambda q}$.
(By ${\rm GL}(2|2)$ invariance, this dimension is independent of the
weight $q$.)  Although $\Delta_\lambda$ cannot be predicted without
knowing the stress-energy tensor of the conformal field theory, we
will make an informed guess later on.  What we can say right away is
that $\Delta_\lambda$ must be an even function of $\lambda$.  (This is
due to the invariance of $\Delta_\lambda$ under a Weyl group action
on the roots of the Lie superalgebra of ${\rm GL}(2|2)$, which takes
$\lambda$ into $-\lambda$.)  The appearance of the weight function
$\mu(\lambda)$ in the denominator, and of the Dirac $\delta$-function
(instead of the usual Kronecker $\delta$-symbol), are forced by the
fact that we are dealing with a continuous series.

On inserting the decompositions (\ref{reduce1}) and (\ref{reduce2})
into the formula (\ref{qmoment}) for $\langle T^q \rangle$, and using
the expression (\ref{twopoint}) for the two-point correlator, we
obtain
\begin{equation}
  \langle T^q \rangle = \int \big| \left\langle V q , V^\ast 0 |
    \lambda q \right\rangle \big|^2 r^{-2\Delta_\lambda} \mu(\lambda)
  d\lambda \;,\label{p-intq}
\end{equation}
where $r$ is the distance between the two point contacts, measured
in the units that are prescribed by the choice of normalization made 
in (\ref{twopoint}).

The computation of the Clebsch-Gordan coefficient and the Plancherel
measure entering in the expression for $\langle T^q \rangle$ is
nontrivial.  Major complications arise from the fact that the modules
$V$ and $V^\ast$ are infinite-dimensional, and that the representation
spaces appearing in the decomposition of the tensor product $V \otimes
V^\ast$ have neither a highest nor a lowest weight vector.  Consequently,
the computation cannot be done solely by algebraic means and a certain
amount of analysis must be invested.  We have relegated this lengthy
calculation to the appendix, where it is shown that
\begin{eqnarray}
  \mu(\lambda)d\lambda &=& {\lambda \over 2} 
  \tanh \left( {\pi\lambda \over 2} \right) d\lambda \;,
\label{plancherel} \\
  \big| \left\langle V q , V^\ast 0 | \lambda q \right\rangle \big|^2 
  &=& {\Gamma(q - {1\over 2} - {i\lambda\over 2})
  \Gamma(q - {1\over 2} + {i\lambda\over 2}) \over \Gamma(q)^2} \;.
\label{CG}
\end{eqnarray}

As it stands, the result for $\langle T^q \rangle$ has been derived
for all positive integers $q$.  However, since the Clebsch-Gordan
coefficient is a meromorphic function of $q$, it is clear that we can
analytically continue the result to all $q$.  (This analytic
continuation is unique, as follows from the bound $|\langle T^q
\rangle| \le 1$ for ${\rm Re}(q) > 0$ and Carlson's theorem stated in
paragraph 5.81 of \cite{titchmarsh}.)  Notice, now, that all of the
$q$-dependence of $\langle T^q \rangle$ resides in the Clebsch-Gordan
coefficient, while the dependence on $r$ is encoded in the factor
$r^{-2\Delta_\lambda}$.  It would therefore seem that all moments of
$T$ decay asymptotically with the same power, $\langle T^q \rangle
\sim r^{-2\Delta_0}$, the exponent being given by $\Delta_0 = {\rm
  Min} _{\lambda \in {\Bbb R}^+} \Delta_\lambda$.  While this is
correct for $q \ge 1/2$, it fails to be true for $q < 1/2$.  The
reason is that, when $q$ is lowered past the value 1/2 from above, the
two poles at $\lambda = \pm i (2q-1)$ of the gamma functions $\Gamma(q
- {1\over 2} \pm {i\lambda \over 2})$ cross the integration axis at
$\lambda = 0$.  Therefore, to do the analytic continuation to $q <
1/2$ correctly, we must pick up the contribution from these poles.  It
is straightforward to calculate the residues at the poles, which
results in
\begin{eqnarray}
  \langle T^q \rangle &=& \Gamma(q)^{-2} \int_0^\infty \big| 
  {\textstyle \Gamma(q - {1\over 2} - {i\lambda\over 2})} \big|^2 
  r^{-2\Delta_\lambda} \mu(\lambda) d\lambda
  \nonumber \\
  &+& 2\pi \cot(q\pi) {\Gamma(2q) \over \Gamma(q)^2} 
  \ r^{-2\Delta_{i(2q-1)}} \quad (|q| < 1/2) \;.
  \label{smallq}
\end{eqnarray}
Here we have used $\Delta_\lambda = \Delta_{-\lambda}$ to combine
terms.

By letting $q$ go to zero in (\ref{smallq}) and using the fact that
normalization of the distribution function for $T$ implies $\langle
T^q \rangle \big|_{q = 0} = 1$, we can deduce a constraint on
$\Delta_\lambda$.  Indeed, from (\ref{CG}) the square of the
Clebsch-Gordan coefficient vanishes uniformly in $\lambda$ as $q \to
0$, and since $\lim_{q\to 0} 2\pi \cot(q\pi) \Gamma(2q) / \Gamma(q)^2
= 1$, we have $\langle T^0 \rangle = r^{-2\Delta_{-i}}$.  Hence we
conclude
  $$
  \Delta_{-i} = \Delta_{+i} = 0 \;.
  $$
Thus, normalization constrains $\Delta_\lambda$ to be of the form
  $$
  \Delta_\lambda = (\lambda^2 + 1) F(\lambda^2) \;.
  $$
It will be convenient to set $F(\lambda^2) = f(\lambda^2 + 1)$.

\subsection{Typical conductance and log-variance}
\label{sec:typical}

Our next goal is to obtain information about the unknown function $f$.
Because we have no analytical control on the fixed point theory, we
are forced to resort to numerical means.  To prepare the numerical
calculation, we first show that $f(0)$ is determined in a very simple
way by how the typical conductance $\exp \langle \ln T \rangle$ varies
with $r$.  To that end we differentiate both sides of (\ref{smallq}) 
with respect to $q$ at $q = 0$ and use the identity
  $$
  {d \over dq} \langle T^q \rangle \big|_{q = 0} 
  = \langle \ln T \rangle \;.
  $$
The integral on the right-hand side of (\ref{smallq}) tends to a
finite value while $\Gamma(q)^{-2}$ behaves as $q^2$ in the limit $q
\to 0$.  Also, the Taylor expansion of $2\pi \cot(q\pi) \Gamma(2q) / 
\Gamma(q)^2 = 1 + {\cal O}(q^2)$ contains no term linear in $q$, as
is easily verified from standard properties of the gamma function.
Hence
  $$
  \langle \ln T \rangle = 
  {d \over dq} r^{-2\Delta_{i(2q-1)}} \big|_{q = 0}
  = - 8 f(0) \ln r \;,
  $$
or, after exponentiation,
\begin{equation}
  \exp \langle \ln T \rangle = r^{-X_{\rm t}} \;, \quad
  X_{\rm t} = 8 f(0) \;. 
\label{typical}
\end{equation}
Thus, $\exp \langle \ln T \rangle$ decays as a pure power.  Note that 
this feature is unique to the typical conductance.  For a general
value of $q$, formula (\ref{smallq}) shows that the behavior of
$\langle T^q \rangle$ as a function of $r$ is governed by a whole {\it
  continuum} of exponents.

We have calculated conductances at criticality ($p = p^\ast = 0.5$)
for systems of size $L = 40$, $60$, and $100$, and for distances $r$
varying from $r = 1$ to $r = L/2$.  For every distance, $2200$
realizations for $L=40$, $1200$ realizations for $L=60$, and between
$200$ and $300$ realizations for $L=100$, were generated.

\begin{figure} [ht]
\leavevmode 
\epsfysize=7cm
\epsfxsize=8cm
\centerline{\epsfbox{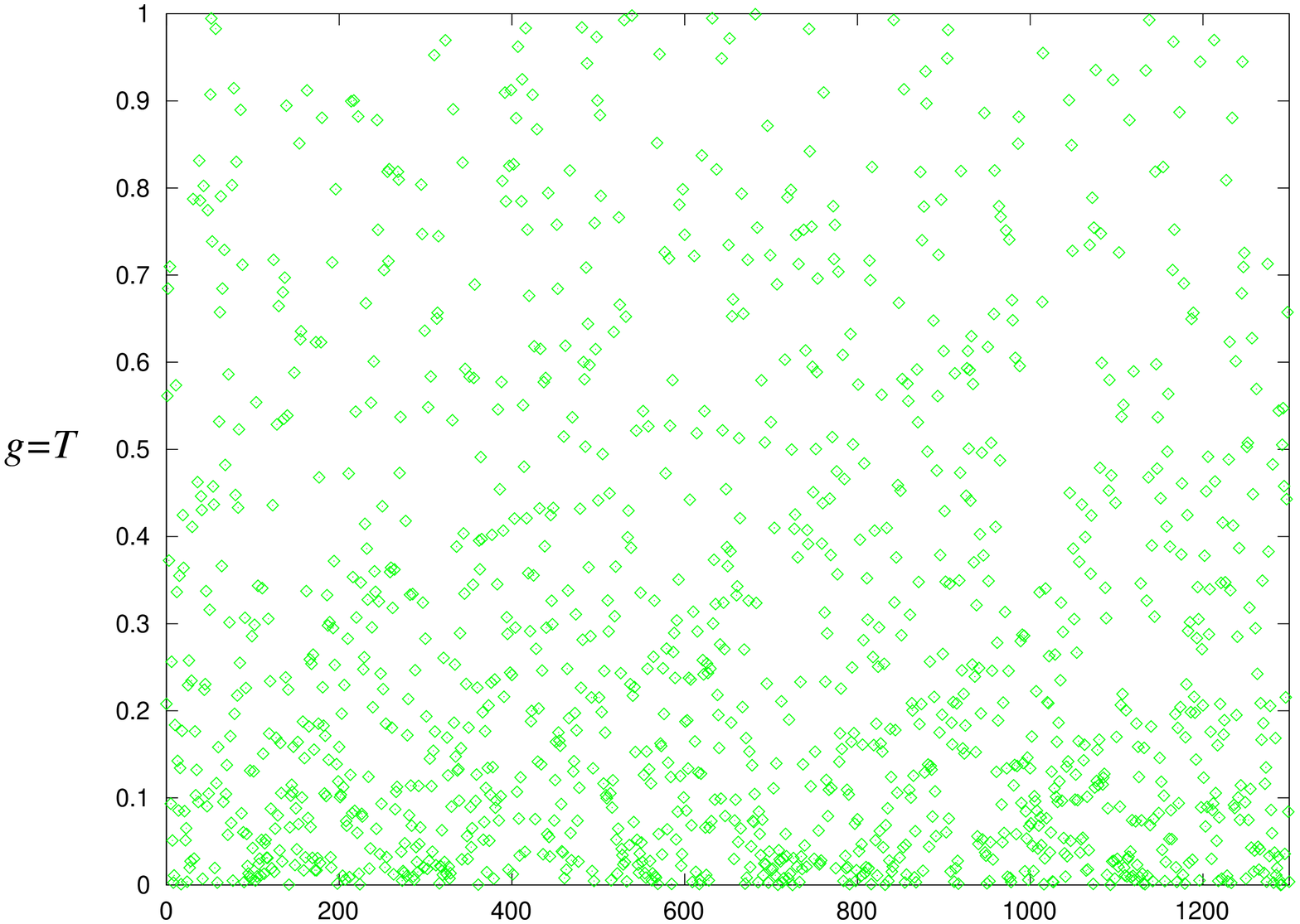}}
\caption{\label{FIG9} Fluctuating values of the conductance $g$ (or
  the transmission probability $T$) at the critical point of the
  network model.  1300 realizations at a fixed distance $r = 10$ are
 displayed.}
\end{figure}

Fig.~{9} shows $1300$ realizations of the conductance for $L
= 60$ and $r = 10$.  The distribution of conductances is seen to be
very broad.  In Fig.~{10} the mean logarithm of the conductance,
$\left\langle \ln T \right \rangle$, is  displayed as a function of 
$\ln r$ for three different system 
sizes $L = 40, 60, 100$.

\begin{figure} [ht]
\leavevmode 
\epsfxsize=7cm
\epsfysize=6cm
\centerline{\epsfbox{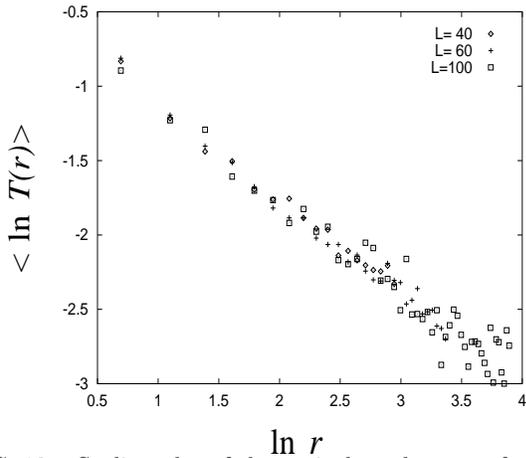}}
\caption{\label{FIG10} Scaling plot of the typical conductance for 
  three different system sizes. The horizontal and vertical axes are
  logarithmic.}
\end{figure}

The typical value indeed scales as a power with $r$,
\begin{eqnarray}
  \exp \langle \ln T \rangle &=&  (r/r_0)^{-X_{\rm t}} 
  \, , \nonumber \\
  X_{\rm t} &=& 0.640 \pm 0.009 \, , \quad
  r_0 \approx  0.42 \, ,
\label{Xt}
\end{eqnarray}
and has no significant dependence on the system size. The result for
$X_{\rm t}$ is based on data for $L=100$ and distances $r\leq 40$, on
data for $L=60$ and distances $r \leq 25$, and on data for $L=40$ and
distances $r \leq 20$.  The error results from a linear fit taking the
statistical errors of the data into account. Taking the same raw data
but neglecting their errors, yields $X_{\rm t}=0.613 \pm 0.012$.  Note
that the hypothesis~(\ref{twopoint}) was formulated in ``conformal
units'' ($r_0 = 1$) whereas the numerical data are represented in
length units given by the lattice constant of the network.  Therefore,
$r_0$ is written explicitly in Eq.~(\ref{Xt}) whereas it is treated as
unity in Eq.~(\ref{typical}).

Higher moments of $\ln T$ can be obtained by using $\langle T^q 
\rangle$ as a generating function:
  $$
  \langle T^q \rangle = 1 + q \langle \ln T \rangle + (q^2/2)
  \langle (\ln T)^2 \rangle + \ldots
  $$
A straightforward calculation starting from (\ref{smallq}) yields
for the log-variance (in conformal units)
\begin{eqnarray}
  \langle (\ln T)^2 \rangle &-& \langle \ln T \rangle^2
  = 16 \big( f(0) - 4 f^\prime (0) \big) \ln r \nonumber\\
  &-& \int_0^\infty ( 1 - r^{-2\Delta_\lambda} ) 
  { 8\pi \ \mu(\lambda) d\lambda \over (\lambda^2 + 1) 
    \cosh(\pi\lambda/2)} \label{log-var}\;.
\end{eqnarray}
The integral on the right-hand side decays algebraically to a constant 
whereas the first term grows logarithmically.  Eq.~(\ref{log-var})
says that the sum, $\Sigma$, of the log-variance and the integral is
linear in $\ln r$.  To obtain the prefactor of $\ln r$, we numerically
calculated the integral for each value of $r$, took the log-variance
from our data for systems of size $L = 60$, and plotted the function
$\Sigma(r)$ versus $X_{\rm t}\ln (r/r_0)$, see Fig.~{11}.
It turned out that the integral was nonnegligible for the range of
$r$ values that were numerically accessible. The slope of the straight
line in Fig.~{11} is $2.08\pm 0.11$, {\it i.e.} the log-variance
is about twice the log-average for large $r$ (as is the case for
in quasi-1D systems \cite{ChaMac}).  Given $8 f(0) = X_{\rm t}$, our 
numerical result implies that $f^\prime(0)$ is very close to zero:
  $$
  -4 f^\prime(0) / X_{\rm t} = 0.005 \pm 0.008 \;,
  $$
which leads us to conjecture $f^\prime = 0$ or, equivalently, a 
dependence of $\Delta_\lambda$ which is exactly quadratic:
\begin{equation}
  \Delta_\lambda = f(0) (\lambda^2 + 1) = {X_{\rm t} \over 8} 
  (\lambda^2 + 1) \;.
  \label{parabolic}
\end{equation}
Aside from being the simplest possible expression for $\Delta_\lambda$,
this guess is in line with field theoretic expectations:  in conformal
field theories with a stress-energy tensor that is quadratic in the
currents, the scaling dimensions are proportional to the quadratic
Casimir invariant.  The polynomial $\lambda^2 + 1$ is, in fact, the
quadratic Casimir of ${\rm GL}(2|2)$, evaluated on the continuous
series of representations $\lambda$.  In the language of multifractality
(Sec.~\ref{sec:multifractal}), the conjecture (\ref{parabolic}) means
that the parabolic approximation to the $F(a)$ spectrum is exact.  In
the remainder of the present paper we shall assume the quadratic form
(\ref{parabolic}). 

\begin{figure} [ht]
\leavevmode 
\epsfxsize=6.5cm
\epsfysize=6.cm
\centerline{\epsfbox{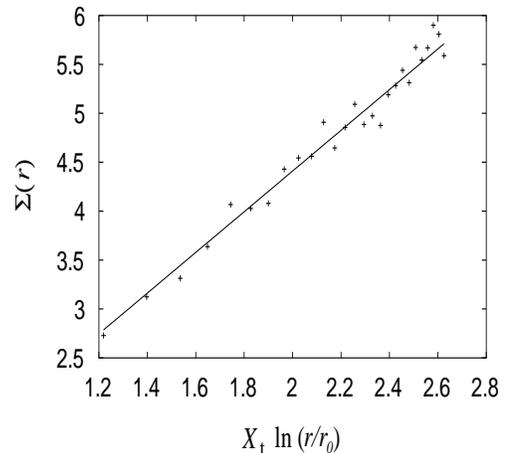}}
\caption{\label{FIG11}  The log-variance of the conductance augmented 
  by the integral on the right-hand side of (\ref{log-var}) is denoted
  by $\Sigma(r)$ and plotted versus $X_{\rm t} \ln(r/r_0)$.}
\end{figure}

\subsection{Multifractal spectrum}
\label{sec:multifractal}

We have seen that the dependence of $\langle T^q \rangle$ on the
distance between the two point contacts is governed in general by a
continuous set of exponents $2\Delta_\lambda$.  This dependence
simplifies, of course, in the asymptotic domain $r \to \infty$.  For
$q \ge 1/2$ the asymptotic behavior is controlled by the smallest
exponent, $2\Delta_0 = X_{\rm t} / 4 \simeq 0.16$.  In the range $-
1/2 \le q \le + 1/2$, the dominant contribution comes from the second
term on the right-hand side of Eq.~(\ref{smallq}), with the exponent
being $2\Delta_{i(2q-1)} = X_{\rm t} q(1-q) \le X_{\rm t}/4$.  For
$q \le -1/2$ there appear additional contributions due to the poles
of $\Gamma(q - {1\over 2} \pm {i\lambda\over 2})$ at $\lambda = \pm
i(2q+1)$, $\pm i(2q+3)$, {\it etc.}  However, since $\Delta_{i(2q-1)}
< \Delta_{i(2q+1)} < \Delta_{i(2q+3)} < \ldots$ for $q < - 1/2$,
these are negligible in the limit $r \to \infty$.  Thus we have
\begin{eqnarray}
  &&\langle T^q \rangle 
  \ {\buildrel r\to\infty \over \sim} \ 
  r^{-X(q)} \;, \nonumber \\
  &&X(q) = \left\{ \matrix{ X_{\rm t}/4 &{\rm for} \ q \ge 1/2 \;, \cr
      X_{\rm t} q(1-q) &{\rm else}\;.\cr} \right.\label{X(q)}
\end{eqnarray}
Note that the spectrum of exponents $X(q)$ is a nondecreasing
function of $q$.  The spectrum shown in Fig.~{12} is nonlinear, 
as is characteristic of a multifractal.

\begin{figure} [ht]
\leavevmode 
\epsfxsize=7.3cm
\epsfysize=5.cm
\centerline{\epsfbox{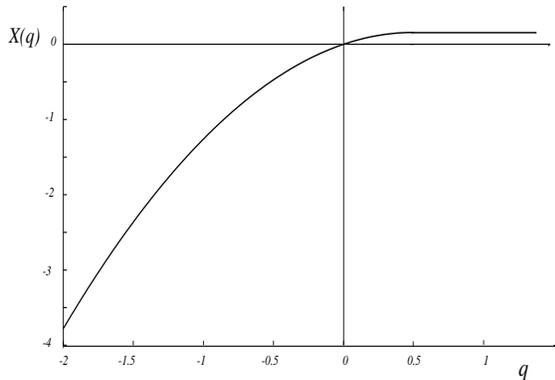}}
\caption{\label{FIG12} The multifractal spectrum $X(q)$ governing 
  the asymptotic power law scaling of the moments $\langle T^q
 \rangle$.}
\end{figure}

That $X(q)$ becomes constant for $q \ge 1/2$ can be traced to the
boundedness of the variable $\ln T \le 0$: for $q \ge 1/2$ the 
asymptotics of the moment $\langle T^q \rangle$ is governed by the
value of the distribution function at the upper bound, ${\rm prob}(T =
1; r) \sim r^{-X_{\rm t}/4}$, which is independent of $q$, as is $T^q
\big|_{T = 1}$. In a thermodynamic interpretation of multifractal
spectra, the nonanalyticity at $q=1/2$ is called a ``phase
transition'' in $X(q)$ \cite{And90}.  The failure of $X(q)$ to become
linear for $q \to -\infty$ results from the absence of a lower bound
on $\ln T$.

The multifractal nature of the $X(q)$ spectrum can alternatively be
described by a spectrum of $f(\alpha)$ type.  If we make a
multifractal ansatz for the probability density of $T$,
  $$
  {\rm prob}(T = r^{-a}) \ {\rm d}T \sim r^{-F(a)} {\rm d}a \;,
  $$
the moments of $T$ for large $r$ scale as
  $$
  \langle T^q \rangle \sim \int r^{-qa - F(a)} da \sim
  r^{-X(q)} \;,
  $$
where $X(q)$ and $F(a)$ are Legendre transforms of each other, {\it 
i.e.} $q + F^\prime(a(q)) = 0$, $X(q) = q a(q) + F(a(q))$, and
conversely, $a - X^\prime(q(a)) = 0$, $F(a) = - a q(a) + X(q(a))$.
Using these relations we find
\begin{equation}
  F(a) = {(a-X_{\rm t})^2 \over 4 X_{\rm t}} \quad (a > 0) \;.
\end{equation}
The $F(a)$ spectrum shown in Fig.~{13} is defined only for 
positive values of $a$ because $a = X^\prime (q) \ge 0$. 

\begin{figure} [ht]
\leavevmode 
\epsfxsize=6.5cm
\epsfysize=6.cm
\centerline{\epsfbox{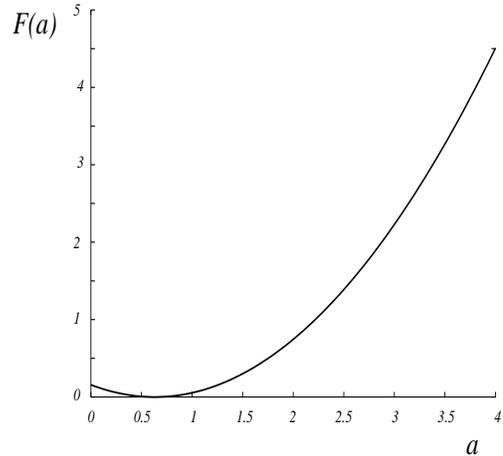}}
\caption{\label{FIG13}  
  The multifractal spectrum $F(a)$ describing the scaling of the
 distribution ${\rm prob}(T=r^{-a}){\rm d}T$.}
\end{figure}

An obvious question arises: is it possible to relate the multifractal
exponents $X(q)$ to the $\tau(q)$ spectrum of the critical
eigenstates, or the scaling exponents of the LDoS?  Our first
observation is that there exists an obvious difference between the two
cases: while the spatial correlations of the LDoS continue to scale in
a nontrivial manner with system size $L$ when $L$ is increased, the
distribution of point-contact conductances (at fixed $r$) becomes {\it
independent} of $L$ \cite{mrz_iqhe}.  Another way of saying this is
that the critical conductance between interior point contacts has a
trivial infinite volume limit, whereas the LDoS does not.

To explain this distinction in physical terms, recall the iterative
procedure by which the stationary limit $\psi_{\infty}$ was
approached.  An essential ingredient in that process was the {\it
  draining action} of the broken links at the two contacts.  By the
loss of probability through these outgoing channels, the system
relaxes and settles down to a stationary long-time limit.  The key to
understanding the size dependence of the conductance is to visualize
the distribution of $|\psi_{\infty}|^2$ in space.  In the regime of
localized states this distribution is concentrated in a circular area
of radius $\sim\xi$ centered around its source ({\it i.e.} the contact
where the current is fed in).  Outside this area the intensity
$|\psi_{\infty}|^2$ falls off exponentially with distance.  On
approaching the critical point, $\xi$ diverges and the exponential
decay turns into a power law.  Thus the intensity becomes more spread
out.  However, in spite of this spreading out, the intensity remains
(algebraically) localized near the source.  The finite size of the
network affects only the tails of the distribution $|\psi_\infty|^2$
and, as a result, the conductance converges to a well-defined limit as
the system size goes to infinity.

When the LDoS is discussed in the same language, a different picture
emerges.  Consider, for simplicity, the $q^{\rm th}$ power of the
density-density correlator
  $$
  \left| \langle c^\prime|(1-{\rm e}^{-\delta}U)^{-1}|c\rangle \right|^{2q}
  $$
which, though not expressible as a correlation function of the LDoS,
is related closely enough to allow a meaningful comparison.  There
again exists a dynamical interpretation.  To compute the
density-density correlator we feed in current through link $c$ and,
after relaxation to the steady state, measure the square of the
amplitude at link $c^\prime$.  What is different from the conductance
measurement is that there are no drains at $c$ and $c^\prime$ (the
network is now isolated).  Instead, particles are absorbed at a
constant rate $2\delta$ at {\it any} location in the network.  As the
system size is increased, the time spent in the network grows as the
Heisenberg time $\delta^{-1} \sim L^d$.  (Previously, the dwell time
in the 2D network was limited by the ``strength'' of the contacts and
remained finite in the limit $L\to\infty$.)  Thus, there is a build up
of particles in the network.  When all states are localized, this
causes the $q$-dependent correlator to diverge as $\delta^{-q} \sim
L^{qd}$.  At the critical point, the divergence from build up is
counteracted by the reduction in the amplitude at link $j$ due to
incipient delocalization of the wavefunctions.  For a general value of
$q$ the two competing effects do not cancel, so that a nontrivial
scaling with $L$ is expected to remain, in agreement with what we
found in Sec.~\ref{sec:LDoS}.

Having said all this, we return to the original question: is it
possible, after all, to relate the multifractal exponents $X(q)$ to
the $\tau(q)$ spectrum?  We wish to offer the following argument.
Imagine placing tunnel barriers at the two contacts $c, c^\prime$.
For zero tunneling probability, the network is closed and we can
measure the density-density correlations.  Now assume that the
$q$-dependent density-density correlator has the same set of scaling
exponents as the LDoS correlator.  Then, in
the limit $\omega = 0$ (or $L_\omega = \infty$) we have
  $$
  \left\langle \Big| \langle c^\prime | (1 - {\rm e}^{-\delta} U)^{-1} | c 
    \rangle \Big|^{2q} \right\rangle \sim r^{-z(q)} L^{z(q)-\tilde z(q)} \;.
  $$
As the tunneling rate is increased, a new time scale appears: the time
a particle injected at link $c$ spends in the network (with the
absorption rate $\delta$ set to zero) before exiting through link
$c^\prime$.  The critical dynamics translates this dwell time into
some characteristic length, $L_\tau$.  Although $L_\tau$ is an
irrelevant length for an almost closed system with high tunnel
barriers, it takes the regularizing role of the system size in the
regime of open systems with $L_\tau < L$. In the limit of vanishing
tunnel barriers, the density-density correlator turns into the
point-contact conductance.  At the same time, the length $L_\tau$ must
become proportional to the distance between the point contacts, for
the simple reason that no other length scale remains available.  This
argument would say $\langle T^q \rangle \sim r^{-\tilde z(q)}$, and
suggests $\tilde z(q)$ as a candidate for $X(q)$.  We should be
cautioned by the fact that the density-density correlator, unlike the
conductance, does not respect any upper bound.  This difference
influences the tails of the distribution and changes the high moments,
at least.  In fact, we have shown $X(q)$ to be constant for $q \ge
1/2$, whereas $\tilde z(q)$ continues to increase.  However, the tails
of the distribution should not affect the typical values, and
therefore one might expect $X_{\rm t} = X^\prime(0) = \tilde z^\prime
(0) = 2(\alpha_0 - 2)$.  Recall that we found $X_{\rm t} = 0.640 \pm
0.009$. Other groups quote values $2(\alpha_0-2) = 0.56 \pm 0.04$
\cite{Pra96}, $0.54 \pm 0.02$ (e.g. \cite{Kle95}), and $0.58 \pm 0.04$
\cite{HuckSchw}.  These values are not all mutually consistent.  We
leave it as an open problem whether there is a flaw in the argument
linking $X(q)$ with $\tilde z(q)$ or there is a real discrepancy.

\subsection{Quasi-1D limit}
\label{sec:quasi1D}

We now endow the network model with a cylinder geometry.  This means
that we consider an infinitely long strip of width $W$, with
coordinates $x \in {\Bbb R}$ and $y \in [0,W]$, and impose periodic
boundary conditions in the transverse direction.  As before, our
interest is in the conductance between two point contacts, which are
placed at positions $(x,y)$ and $(x',y')$.  The conductance in such a
cylindrical setting can be related to the point-contact conductance in
the infinite 2D plane with complex coordinate $z$ by a conformal
transformation
  $$
  z = \exp \ {2\pi \over W}(x+iy) \;.
  $$
The conformal field theory rule for translating two-point functions
from the plane to the cylinder reads \cite{cardy}
\begin{eqnarray*}
  &&\langle \varphi_\lambda(z) \varphi_\lambda(z^\prime) \rangle = 
  |z - z^\prime|^{-2\Delta_\lambda} \\
  &&\to \left| {W \over \pi} \sinh \left( {\pi\over W}(x-x^\prime 
      +iy-iy^\prime) \right) \right|^{-2\Delta_\lambda} \;.
\end{eqnarray*}
From this rule we have the relation
  $$
  \langle T^q \rangle \big|_{(x,y);(x',y')}^{\rm cylinder} = 
  \langle T^q \rangle \big|_{r = | {W \over \pi} \sinh ( {\pi\over W} 
    (x-x'+iy-iy')) |}^{\rm 2D} \;.
  $$
In particular, for the typical cylindrical conductance we obtain
  $$
  \exp \langle \ln T \rangle = 
  \left| {W \over \pi} \sinh \left( {\pi\over W}(x-x^\prime + iy
      -iy^\prime) \right) \right|^{-X_{\rm t}} \;.
  $$
(Recall that we are using length units so that $\langle \ln T \rangle_
{2{\rm D}} = 0$ for $r = 1$.) In the quasi-1D limit $L \equiv |x - x^
\prime | \gg W$, this result simplifies to
  $$
  \exp \langle \ln T \rangle \big|_{L \gg W} = (W / 2\pi)^{X_{\rm t}}
  \exp - \pi X_{\rm t} L / W \;.
  $$
Note that from (\ref{Xt}) the numerical value of $\pi X_{\rm t}$ is
\begin{equation}
  \pi X_{\rm t} = 2.01 \pm 0.03 \;.
\label{puzzle}
\end{equation}

What is this result, a value of $\pi X_{\rm t}$ close to 2, trying to 
tell us? Let us offer some speculation based on the assumption that
the relation $\pi X_{\rm t} = 2$ holds exactly.  In \cite{mrz_network}
it was shown that, if a naive continuum limit is assumed ({\it i.e.}
possible renormalization effects due to short wave length modes are
ignored), the supersymmetric vertex model (\ref{latticeFT}) for the
critical network is equivalent to Pruisken's nonlinear $\sigma$ model
\cite{Pru84} at couplings $\sigma_{xx} = \sigma_{xy} = 1/2$. The
action functional of the latter model is
  $$
  {\cal S} = \int dx dy \ \Big( \sigma_{xx} ( {\cal L}_{xx} + 
    {\cal L}_{yy}) + \sigma_{xy} ( {\cal L}_{xy} - {\cal L}_{yx} )
    \Big) \;,
  $$
where
  $$
  {\cal L}_{\mu\nu} = {\rm STr} (1 - Z\tilde Z)^{-1} \partial_\mu Z
  (1 - \tilde Z Z)^{-1} \partial_\nu \tilde Z \;.
  $$
For the sake of the argument, let us now {\it assume} Pruis\-ken's model 
at $\sigma_{xx} = \sigma_{xy} = 1/2$ to be a {\it fixed point} of the 
renormalization group.  Then, by raising the short distance cutoff
from $a = 1$ to $a = W$ we can reduce Pruisken's action ${\cal S}$
to a 1D effective action
  $$
  {\cal S}_{1{\rm D}} = {W \over 2} \int dx \ {\rm STr} 
  (1 - Z\tilde Z)^{-1} \partial_x Z
  (1 - \tilde Z Z)^{-1} \partial_x \tilde Z \;.
  $$
Alternatively, we could argue that in the quasi-1D limit the 
dependence of the field $Z$ on $y$ can be neglected and, since
$\sigma_{xx}$ does not renormalize (by the fixed point hypothesis),
the process of scaling out $y$ simply produces a factor $\int dy = W$.

The one-dimensional theory with action functional ${\cal S}_{1{\rm
    D}}$ has been much studied, and its mean conductance is known
\cite{Efe82,Zir92,mmz} to decay with length $L$ of the conductor,
which plays the role of distance between the contacts, as
  $$
  \langle T \rangle \big|_{L}^{\rm 1D} \sim \exp - L / 2W \;.
  $$
Moreover, from \cite{mf} we know that the localization lengths for
the mean and typical conductances of the one-dimensional nonlinear
$\sigma$ model differ by a factor of 4, so that
  $$
  \exp \langle \ln T \rangle \big|_{L}^{\rm 1D} \sim \exp - 2L / W \;,
  $$
which agrees, within the numerical errors, with what we found in
(\ref{puzzle}).  To turn the argument around, by assuming Pruisken's
nonlinear $\sigma$ model at $\sigma_{xx} = \sigma_{xy} = 1/2$ to be a
fixed point, we would have predicted $X_{\rm t}$ to be
  $$
  X_{\rm t} = (\pi\sigma_{xx})^{-1} \big|_{\sigma_{xx} = 1/2} 
  \approx 0.637 \;.
  $$

The above argument is not convincing, as it relies on the questionable
assumption that the nonlinear $\sigma$ model is a fixed point theory.
Conventional wisdom has it that critical two-dimensional nonlinear
$\sigma$ models are unstable with respect to quantum fluctuations and
flow under renormalization to theories of the Wess-Zumino-Witten type.
However, we can reformulate the argument and avoid any reference to
Pruisken's theory.  Let us assume that it is the network model itself
(or, rather, a suitable continuum limit thereof) which is a fixed
point of the renormalization group.  Note that such an assumption is
consistent with the fact that network model observables start scaling
very rapidly when the observation scale is increased. (For example, in
Fig.~{10} there are no visible deviations from linearity as $r$
approaches the short distance cutoff $a$.)  As before, we imagine
raising the cutoff by using a sequence of RG transformations.  By the
fixed point hypothesis, we arrive for $a = W$ at the 1D network model
(or, rather, some continuum version closely related to it), with the
distance between contacts rescaled to $L / W$.  Next we pass to the 1D
supersymmetric vertex model, and from there to the continuum action
${\cal S}_{1{\rm D}}$.  In contrast to earlier, the last step is
benign, as the 1D nonlinear $\sigma$ model is {\it
  superrenormalizable} ({\it i.e.} ultraviolet finite) and the RG
trajectory can no longer depart from it.  By this token, we again
arrive at $X_{\rm t} = 2/\pi$, this time {\it without} having passed
through Pruisken's model.  Thus, the proposed value for $X_{\rm t}$
follows as a consequence of assuming the network model (or a suitable
continuum limit thereof) to be a RG fixed point.

Note that if the above fixed point assumption is correct and $X_{\rm
  t} = 2/\pi$ holds {\it exactly}, then we are led to the striking
conclusion that Wess-Zumino-Witten models are ruled out as candidates
for the fixed point theory.  Indeed, the scaling dimensions for such
models are given \cite{KZ} by ${\cal C}_\lambda / (k+h_\ast)$, where
${\cal C}_ \lambda$ is the quadratic Casimir, and $k$ and $h_\ast$ are
integers. Such an expression for the scaling dimensions cannot produce 
the irrational number $2/\pi$.

\subsection{Reconstruction of the distribution}
\label{sec:distribution}

We now return to the two-dimensional network and reconstruct the
entire distribution function for the point-contact conductance from
the moments $\langle T^q \rangle$.  By making a simple variable
substitution (namely $x = 2\rho + 1$) in Eq.~(\ref{intrep}) of the
appendix, the product of gamma functions in the formula for $\langle
T^q \rangle$ can be represented as an integral over Legendre
functions:
  $$ 
  { |\Gamma(q - {1\over 2}-{i\lambda\over 2})|^2 \over \Gamma(q)^2} =
  \int_0^\infty (1+\rho)^{-q} {\cal P}_{(i\lambda-1)/2}(2\rho+1) d\rho \;.
  $$
Next we define a probability density ${\rm prob}(\rho;r) {\rm d}\rho$ 
for the variable $\rho$ by
  $$
  {\rm prob}(\rho;r) = \int_0^\infty r^{-X_{\rm t}(\lambda^2 + 1)/4}
  {\cal P}_{(i\lambda-1)/2}(2\rho + 1) \mu(\lambda) d\lambda \;.
  $$
Given $2\Delta_\lambda = X_{\rm t} (\lambda^2+1)/4$, comparison with
Eqs.~(\ref{p-intq}) and (\ref{CG}) yields
  $$
  \langle T^q \rangle = \int_0^\infty (1+\rho)^{-q} 
  {\rm prob}(\rho;r) d\rho \;.
  $$
Hence, on making the identification $T \equiv (1 + \rho)^{-1}$ we
conclude that the probability density for $T$ is ${\rm prob}(\rho;r)
{\rm d}\rho$.  Although this is easily expressed in terms of $T$ by
using the inverse relation $\rho = T^{-1} - 1$, which has differential
${\rm d}\rho = - T^{-2} {\rm d}T$, we find it more convenient to work
with the variable $\rho$ instead of $T$.

Because the Legendre functions ${\cal P}_{(i\lambda-1)/2}(2\rho+1)$
are oscillatory with respect to $\lambda$ (incidentally, they
oscillate also w.r.t.~$\rho$), the above formula for the probability
density is not well suited for numerical evaluation.  Motivated by
this, we switch to a different representation as follows.  The
Legendre functions satisfy the hypergeometric differential equation
  $$
  \left( {1\over 4}(\lambda^2+1) + {\partial \over \partial\rho} 
    \rho (\rho + 1) {\partial \over \partial \rho} \right) 
  {\cal P}_{(i\lambda-1)/2}(2\rho + 1) = 0 \;,
  $$
and the integral of ${\cal P}_{(i\lambda-1)/2}$ against the Plancherel
measure gives a Fourier representation of the $\delta$-function:
  $$
  \delta(\rho) = \int_0^\infty {\cal P}_{(i\lambda-1)/2}(2\rho+1)
  \mu(\lambda)d\lambda \;.
  $$
Both facts are standard results in harmonic analysis on the hyperbolic
plane (or Lobachevsky plane) and are briefly reviewed in the appendix.
Using them in the $\lambda$-integral representation for ${\rm prob}
(\rho;r)$ we obtain
\begin{eqnarray}
  \left( {\partial \over \partial \ln r} + X_{\rm t} 
    {\partial \over \partial\rho} \rho (\rho + 1) {\partial \over
      \partial\rho} \right) {\rm prob}(\rho;r) &=& 0 \;, \nonumber \\
  \lim_{\ln r\to 0} {\rm prob}(\rho;r) &=& \delta(\rho) \;.
\label{heateqn}
\end{eqnarray}
Consider now the hyperbolic plane with the metric tensor in polar
coordinates $\theta,\phi$ given by ${\rm d}\theta^2 + \sinh^2(2\theta)
{\rm d}\phi^2$.  If we substitute $\rho = \sinh^2 \theta$, the
differential operator $\partial_\rho \rho(\rho+1) \partial_\rho$
turns into 
  $$
  {\partial\over\partial\rho} \rho(\rho+1)
  {\partial\over\partial\rho} = 
  {1 \over 4} {1 \over \sinh(2\theta)} {\partial\over\partial\theta}
  \sinh(2\theta) {\partial\over\partial\theta} \;,
  $$
which coincides with 1/4 times the radial part of the Lap\-lace-Beltrami
operator on the hyperbolic plane.  Therefore, by viewing $\ln r$ as
``time'' and $X_{\rm t}/4$ as a ``diffusion constant'', we can
interpret the initial value problem (\ref{heateqn}) as the heat (or
diffusion) equation on that space.  Solving the heat equation on the
hyperbolic plane is a textbook example in Riemannian geometry 
\cite{heatkernel}.  For our purposes, a convenient expression for the 
solution is the following integral:
  $$
  {\rm prob}(\rho;r) = { 2 \pi^{-1/2} \ r^{-X_{\rm t}/4} \over 
    (X_{\rm t} \ln r)^{3/2} } \int\limits_{{\rm arcsinh}
    \sqrt{\rho}}^\infty {{\rm e}^{-t^2 / (X_{\rm t} \ln r)} t dt 
    \over \sqrt{\sinh^2(t) - \rho}} \;,
  $$
which is easy to compute numerically.  The result for the distribution 
function 
\begin{equation}
  f(\ln T;r) = {\rm prob}(\rho;r/r_0) \left| {d\rho \over d\ln T} \right|
  \label{distrib}
\end{equation}
is plotted in Fig.~{14} for the distance $r=15$ between the
contacts.  The value $X_{\rm t} = 0.64$ is assumed.  The error bars
correspond to the mean deviation to be expected in histograms
accumulated from 1760 independent measurements of data following the
predicted distribution. It is seen that our analytical prediction
agrees well with the numerical data points (accumulated from 1760
conductances) represented by dots.

\begin{figure} [ht] 
\leavevmode 
\epsfysize=6cm
\centerline{\epsfbox{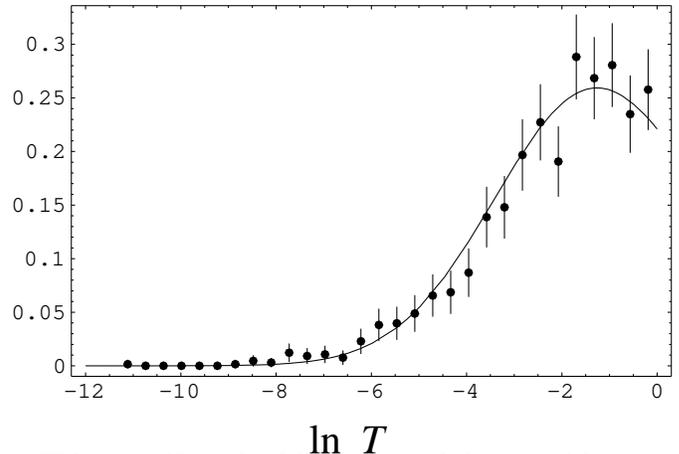}}
\caption{\label{FIG14} Normalized histograms of the critical 
  log-con\-ductance $\ln T$ (dots) for distance $r = 15$.  
The numerical data are compared with plots of the
  distribution function $f(\ln T;r)$ defined in (\ref{distrib}).}
\end{figure}

\section{Summary}

We have presented a numerical and analytical study of point-contact
conductance distributions for the Chalker-Coddington network model.
After reconsidering the multifractal correlations of the local density
of states, we first focussed on the distribution of point-contact
conductances in the quantum Hall plateau region, where strong
localization of electrons occurs. As expected, the distribution is
close to log-normal and is essentially parameterized by only the
typical localization length. In particular, we found the log-variance
to be proportional to the logarithm of the typical conductance,
with the constant of proportionality being $-1.00 \pm 0.05$.

We then turned to the plateau-to-plateau transition of the quantum
Hall effect.  Our analytical results are summarized as follows.  By
transforming the network model to a supersymmetric vertex model with
${\rm GL}(2|2)$ symmetry, we derived a formula, Eq.~(\ref{p-intq}),
for the $q^{\rm th}$ moment of the point-contact conductance at
criticality.  The general structure of the formula is completely
determined by group symmetry.  The unknowns are the scaling dimensions
$\Delta_ \lambda$ of certain local operators $\varphi_{\lambda q}$,
which represent the point contacts in the formulation by the vertex
model.  We assumed these scaling dimensions to be proportional to the
quadratic Casimir invariant of the symmetry algebra: $\Delta_ \lambda
= X_{\rm t}(\lambda^2+1)/8$.  (This assumption is not essential and
can in principle be relaxed.)  This choice leaves $X_{\rm t}$ as the
only free parameter.  Knowledge of all the moments allowed us to
reconstruct the entire distribution function.

Salient predictions of our analysis are: i) The distribution of
point-contact conductances becomes independent of the system size $L$
in the thermodynamic limit $L \to \infty$.  ii) At the critical point,
the typical point-contact conductance of the infinite 2D network
decays with the distance $r$ between the two contacts as a pure power:
$\exp \langle \ln T \rangle = (r/r_0)^{-X_{\rm t}}$.  iii) The
log-variance equals $-2$ times the logarithm of the typical
conductance.  iv) For large distances between the contacts, the
$q$-moments of the conductance exhibit multifractal statistics:
$\langle T^q \rangle \sim r^{-X(q)}$, where $X(q) = -X_{\rm t} q
(q-1)$ for $q \le 1/2$ and $X(q) = X_{\rm t} / 4$ for $q \ge 1/2$.
Thus there is a ``phase transition'' in the $X(q)$ spectrum at $q =
1/2$.

All these predictions are consistent with our numerical data, which
were accumulated by a computing effort of about $2000$ CPU hours on a
Sun Sparc workstation.  We found the distribution of point-contact
conductances for $r < L/2$ to show no significant dependence on the
system size, as expected.  In a double logarithmic plot of the typical
conductance versus $r$, the data points scatter around a linear curve
with slope $-X_{\rm t} = - 0.640 \pm  0.009$.  The log-variance is
linearly related to the logarithm of the typical conductance, with the
constant of proportionality being $-2.08 \pm 0.11$. The phase
transition in the $X(q)$ spectrum is hard to see in our data, since
the numerically accessible values of $r$ are not large enough in order
for the asymptotic behavior to dominate.  However, the predicted
distribution function for the point-contact conductances agrees well
with our numerical data.

On a speculative note we argued that, if the network model (or,
rather, a suitable continuum limit thereof) is a fixed point of the
renormalization group, then the scaling exponent for the typical
point-contact conductance must have the value $X_{\rm t} = 2/\pi$.
This follows from conformal invariance linking 2D with quasi-1D, and
from exact results available for the latter.  While the fixed point
assumption for the network model needs to be substantiated, it is
remarkable that the predicted value lies very close to the numerical
result.

As a suggestion for further work, recall from Sec.~\ref{sec:quasi1D}
that conformal invariance at the critical point predicts the typical
conductance between two point contacts at positions $(0,0)$ and
$(x,y)$ on a cylinder with circumference $W$ to be
  $$
  \exp \langle \ln T \rangle_{(x,y);(0,0)}^{\rm cylinder} = 
  \left| {W \over \pi} \sinh \left( {\pi \over W} (x + iy) \right)
  \right|^{-X_{\rm t}} \;.
  $$
Verification of this relation would provide a stringent test of the
idea of a conformal fixed point theoy for the quantum Hall transition.
We have not done the test, as our numerical calculations had already
been long completed by the time we became aware of the exactness of
the relation.  We invite other groups to perform this stringent test
and reduce the statistical error on $X_{\rm t}$.  We feel certain that
the value of $X_{\rm t}$ will be a benchmark for the analytical theory
yet to be constructed, and is desirable to know with the same accuracy
as the localization length exponent $\nu$.  \smallskip

{\bf Acknowledgment.}  We thank Alexander Altland for reading the manuscript.
This research was supported in part by the
Sonderforschungsbereich 341 (K\"oln-Aachen-J\"ulich).

\appendix
\section{}

Here we derive the result for the Clebsch-Gordan coefficient and the
Plancherel measure announced in equations (\ref{CG}) and
(\ref{plancherel}).  For reasons that were explained in the text, this
isn't an easy calculation.  Fortunately, we can do it by using the
following trick.
 
We consider a very simple network, consisting of just two edges that
interact along a chain of $L$ vertices (see Fig.~15).

\begin{figure} [ht]
\leavevmode
\epsfxsize=7cm
\centerline{\epsfbox{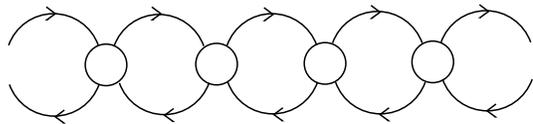}}
\caption{\label{FIG15}  Graphical representation of a network,
consisting of two edges that interact along a chain of vertices.}
\end{figure}

  The
formalism developed in Secs. \ref{sec:mapping} and
\ref{sec:symmetries} applies to this case just as well as to the
two-dimensional network model.  In particular, the $q^{\rm th}$ moment
of the conductance is given by a formula such as (\ref{qmoment}).  A
simplifying feature is that the product of ${\cal R}$ matrices now
organizes into a convolution product of transfer matrices, ${\cal
  T}^L$.  Denoting the eigenvalues of the transfer matrix ${\cal T}$
by $t(\lambda)$ we get
\begin{equation}
  \langle T^q \rangle_{1{\rm D};L} = 
  \int_0^\infty \big| \left\langle V q , V^\ast 0 | \lambda q 
  \right\rangle \big|^2 t(\lambda)^L \mu(\lambda) d\lambda \;,
\label{1Dmoments}
\end{equation}
by a similar reasoning as in the body of the paper.  Our strategy will
now be to exploit the simplicity of this 1D model and compute the
moments $\langle T^q \rangle_{1{\rm D};L}$ from a quite different
approach.  By comparing the result to the formula (\ref{1Dmoments}),
we will ultimately be able to read off the desired expressions for the
Clebsch-Gordan coefficient and the Plancherel measure.

For technical convenience, we shall consider the two-edge network
model in the limit of weak backscattering at the nodes.  An attractive
feature of this limit is that the task of computing $\langle T^q
\rangle_{1{\rm D};L}$ can be reformulated as a {\it partial
  differential equation} (of the Fokker-Planck type) which is readily
solved.  Before writing down that equation, it is helpful to make two
adjustments.  Rather than computing directly the moments $\langle T^q
\rangle$, we will study the {\it entire distribution function} of $T$.
Also, we switch from $T$ to the variable $\rho=(1-T)/T$ (``Landauer's
resistance'').  Now, by an elementary calculation (for a review, see
{\it e.g.} \cite{beenakker}), the probability density ${\rm prob}
(\rho;L) {\rm d}\rho$ of Landauer's resistance satisfies the
differential equation
\begin{equation}
  \ell {\partial \over \partial L} {\rm prob}(\rho;L) = 
  {\partial \over \partial\rho} \rho (\rho + 1) {\partial \over
    \partial \rho} {\rm prob}(\rho;L)
\label{fp}
\end{equation}
where $\ell$ is the elastic mean free path.  We are going to solve
this equation by harmonic analysis, {\it i.e.} by diagonalization of
the differential operator $\partial_\rho \rho (\rho + 1) \partial_
\rho$.  [This operator has a geometric meaning as the radial part of
the Laplacian on a noncompact Riemannian symmetric space ${\rm
  SU}(1,1)/{\rm U}(1)$.]  Introducing the Legendre function ${\cal
  P}_\nu (x)$ through its integral representation,
\begin{equation}
  {\cal P}_\nu (x) = \int_0^{2\pi} {d\phi \over 2\pi} 
  \left( x + \sqrt{x^2 - 1} \cos\phi \right)^\nu \;,
\label{legendre}
\end{equation}
one easily verifies
  $$
  \left( {\partial \over \partial\rho} \rho (\rho + 1) {\partial \over
      \partial \rho} + {1 \over 4} (\lambda^2 + 1) \right) 
  {\cal P}_{(i\lambda-1)/2}(2\rho + 1) = 0 \;.
  $$
This relation suggests a solution of the differential equation
(\ref{fp}) of the form
  $$
  {\rm prob}(\rho;L) = \int_0^\infty {\rm e}^{-{L \over 4\ell}
    (\lambda^2 + 1)} {\cal P}_{(i\lambda-1)/2}(2\rho + 1) 
  d m(\lambda) \;.
  $$
The spectral measure (or Plancherel measure) $d m(\lambda)$ is
determined by the asymptotic behavior of the Legendre functions ${\cal
  P}_{(i\lambda-1)/2}(2\rho + 1)$ for $\rho \to \infty$, as follows.
By using the substitution $u = \tan (\phi/2)$ in the integral
representation (\ref{legendre}), one finds
  $$
  {\cal P}_{(i\lambda-1)/2} (\cosh 2\theta) \;
  {\buildrel \theta \to \infty \over \longrightarrow} \; 
  {\rm e}^{-\theta} \left( {\bf c}(\lambda) {\rm e}^{i\lambda\theta} 
    + {\bf c}(-\lambda) {\rm e}^{-i\lambda\theta} \right)
  $$
where the {\bf c}-function is given by
  $$
  {\bf c}(\lambda) = {1 \over \sqrt{\pi}} { \Gamma({1\over 2}i\lambda)
    \over \Gamma({1\over 2}(i\lambda + 1))} \;.
  $$
{}From this asymptotic limit, we infer the orthogonality relations
  $$
  \int_0^\infty {\cal P}_{(i\lambda - 1)/2}(2\rho + 1)
  {\cal P}_{(i\lambda^\prime - 1)/2}(2\rho + 1) d\rho = 
  N_\lambda \delta(\lambda-\lambda^\prime) \;,
  $$
with the normalization factor being $N_\lambda = \pi |{\bf c}(\lambda)
|^2$.  In conjunction with the initial condition
  $$
  \lim_{L \to 0+} {\rm prob}(\rho;L) = \delta(\rho) \;,
  $$
stating that transmission through a short chain is ideal, the presence
of the normalization factor determines the spectral measure to be
  $$
  d m(\lambda) = {{\rm d}\lambda \over \pi |{\bf c}(\lambda)|^2}
  = {\rm \lambda \over 2} \tanh \left( {\pi\lambda \over 2} \right)
  {\rm d}\lambda \;.
  $$

To recover the moments $\langle T^q \rangle$ from the distribution
${\rm prob}(\rho;L) {\rm d}\rho$, we need the integral
  $$
  I_\lambda(q) \equiv \int_0^\infty (1 + \rho)^{-q} 
  {\cal P}_{(i\lambda - 1)/2} (2\rho + 1) d\rho \;,
  $$
which converges for $q > 1/2$.  We claim that this integral has the 
value
  $$
  I_\lambda(q) = { \Gamma(q - {1 \over 2} + {i\lambda \over 2})
    \Gamma(q - {1 \over 2} - {i\lambda \over 2}) \over \Gamma(q)^2}
  \;.
  $$
To prove this statement, we proceed as follows.  In the first step, 
we set $x = 2\rho + 1$ and write
\begin{eqnarray*}
  &&\int_1^\infty (x+1)^{-q} {\cal P}_{(i\lambda-1)/2} (x) dx \\
  &=& \Gamma(q)^{-1} \int_0^\infty \left( \int_1^\infty {\rm e}^{-ax}
    {\cal P}_{(i\lambda - 1)/2}(x) dx \right) a^{q-1} {\rm e}^{-a} da
  \;, 
\end{eqnarray*}
which is a valid equality for $q > 0$.  According to Ref.~\cite{erdelyi} 
(p. 323, no. 11) the integral in parentheses equals
  $$
  \int_1^\infty {\rm e}^{-ax} {\cal P}_{(i\lambda-1)/2}(x) dx
  = \sqrt{2 \over \pi a} K_{i\lambda/2}(a) \;.
  $$
This leads to an integral over the auxiliary variable $a$, which 
converges for $q > 1/2$ and the value of which we take from 
Ref.~\cite{gradzhteyn} [p. 716, no. 6.628(7)]:
\begin{eqnarray*}
  &&\sqrt{2 \over \pi} \int_0^\infty a^{q-3/2} {\rm e}^{-a} 
  K_{i\lambda/2}(a) da = \\
  &&\Gamma(q - {1\over 2} - {i\lambda\over 2})
  \Gamma(q - {1\over 2} + {i\lambda\over 2})
  \lim_{\alpha\to 0} { {\cal P}_{(i\lambda-1)/2}^{1-q}(\cosh\alpha)
    \over (\sinh\alpha)^{q-1} } \;,
\end{eqnarray*}
where ${\cal P}_\nu^\mu(x)$ is the associated Legendre function. 
{}From Ref.~\cite{abramowicz} (p. 332, no. 8.1.2) this function has
the following small-$\alpha$ limit:
  $$
  {\cal P}_{(i\lambda-1)/2}^{1-q}(\cosh\alpha) \;
  {\buildrel \alpha\to 0 \over \longrightarrow} \;
  \Gamma(q)^{-1} (\alpha/2)^{q-1} \;.
  $$
Combination of all these results yields
\begin{eqnarray}
  I_\lambda(q) &=& 2^{q-1} \int_1^\infty (1+x)^{-q} 
  {\cal P}_{(i\lambda-1)/2}(x) dx \nonumber \\
  &=& { \Gamma(q - {1\over 2}-{i\lambda\over 2})
  \Gamma(q - {1\over 2}+{i\lambda\over 2}) \over \Gamma(q)^2} \;,
\label{intrep}
\end{eqnarray}
which proves the claim.

By substituting $T = (1+\rho)^{-1}$ and inserting for the probability
density ${\rm prob}(\rho;L) {\rm d}\rho$ the spectral resolution given
earlier, we finally arrive at
  $$
  \langle T^q \rangle_{1{\rm D};L} = \int_0^\infty {\rm e}^{
    - (\lambda^2 + 1) L / 4\ell} I_\lambda(q) dm(\lambda) \;.
  $$
Comparison with (\ref{1Dmoments}) identifies the eigenvalue of the
transfer matrix ${\cal T}$ as $t(\lambda) = {\rm e}^{-(\lambda^2+1)/4\ell}$,
and yields
\begin{equation}
  \big| \langle V q , V^\ast 0 | \lambda q \rangle \big|^2 
  \mu(\lambda) d\lambda = I_\lambda(q) dm(\lambda) \;.
\label{cgp}
\end{equation}
Although this result gives an answer for the {\it product} of the 
squared Clebsch-Gordan coefficient with the Plancherel measure, it
does not allow to make the separate identifications proposed in
Eqs.~(\ref{CG}) and (\ref{plancherel}).  For that, more detailed
considerations are necessary.  For brevity, we refrain from
elaborating on these since, actually, all that is needed for the main
text is the formula (\ref{cgp}).

\end{document}